\documentclass[11pt,journal,onecolumn,draftcls,english]{IEEEtran}

\usepackage[]{fontenc}
\usepackage[latin1]{inputenc}
\usepackage{verbatim}
\usepackage{amsmath}
\usepackage{graphicx}
\usepackage{amssymb}
\usepackage{color}
\usepackage{ulem}
\usepackage{xspace,bbm,epic,eepic}
\usepackage[noadjust]{cite} 
\usepackage{theorem}

\title{Exact Recovery Conditions for Sparse Representations\\ with Partial Support Information}

\author{C.~Herzet$^\star$,~C.~Soussen,~J.~Idier,~and~R.~Gribonval
 \thanks{C.~Herzet and R.~Gribonval are with INRIA Rennes - Bretagne
    Atlantique, Campus de Beaulieu, F-35042 Rennes Cedex, France
    (e-mail: Cedric.Herzet@inria.fr; Remi.Gribonval@inria.fr).}
 \thanks{C.~Soussen is with the \CRAN (\cran). Campus Sciences, B.P.
    70239, F-54506 Vand{\oe}uvre-l\`es-Nancy, France (e-mail:
    Charles.Soussen@univ-lorraine.fr.)}% <-this % stops a space
  \thanks{J.~Idier is with the Institut de Recherche en Communications
    et Cybern\'etique de Nantes (IRCCyN, UMR CNRS 6597), BP 92101, 1 rue
    de la No\"e, 44321 Nantes Cedex~3, France (e-mail:
    Jerome.Idier@irccyn.ec-nantes.fr).}
}%\name {XXX and YYY}

\def\XS{\xspace}
\DeclareMathAlphabet{\mathb}{OML}{cmm}{b}{it}
\def\tb{{\sbm{t}}\XS}

\RequirePackage{ifthen}
\RequirePackage{calc}

\newcommand{\taille}[1][\scad]{%
\ifthenelse{#1 = -5}{}{}%
\ifthenelse{#1 = -4}{\tiny}{}%
\ifthenelse{#1 = -3}{\scriptsize}{}%
\ifthenelse{#1 = -2}{\footnotesize}{}%
\ifthenelse{#1 = -1}{\small}{}%
\ifthenelse{#1 = 0}{\normalsize}{}%
\ifthenelse{#1 = 1}{\large}{}%
\ifthenelse{#1 = 2}{\Large}{}%
\ifthenelse{#1 = 3}{\LARGE}{}%
\ifthenelse{#1 = 4}{\huge}{}%
\ifthenelse{#1 = 5}{\Huge}{}}

\newboolean{@serif}
\setboolean{@serif}{true}
\def\scad{-5} % scadefaultsize

\newcounter{taille}
\newcommand{\sca}[2][\scad]{\setcounter{taille}{#1}%
  \ifthenelse{\boolean{@serif}}
  {{\taille[\thetaille]\textsc{#2}}}
  {\setcounter{taille}{\value{taille}-1}{\uppercase{\taille[\thetaille]#2}}}}

\def\eg{\textit{e.g.,}\XS}
\def\etal{\textit{et al.}\XS}

\def\ie{\textit{i.e.,}\XS}

\def\CRAN{Centre de Recherche en Automatique de Nancy\XS}
\newcommand{\cran}[1][\scad]{\sca[#1]{cran, umr 7039},
   Universit\'e de Lorraine, \sca[#1]{cnrs}\XS}

\newcommand{\adresseCRAN}[1][\scad]{Campus Sciences, B.P. 70239, 
F-5   4506 Vand{\oe}uvre-l\`es-Nancy, France\XS}

                \def\stdpth#1{(#1)}
              \def\stdacc#1{\{#1\}}

   \def\stdscal#1{\langle#1\rangle}

\def\spansub#1{{\mathrm{span}}\stdpth{#1}}

\def\spark#1{{\mathrm{spark}}\stdpth{#1}}

\def\a{{\mathbf a}}
\def\b{{\mathbf b}}

\def\x{{\mathbf x}}
\def\y{{\mathbf y}}

\def\v{{\mathbf v}}
\def\w{{\mathbf w}}

\def\r{{\mathbf r}}
\def\u{{\mathbf u}}

\def\ta{{\tilde{\mathbf a}}}
\def\tb{{\tilde{\mathbf b}}}
\def\tc{{\tilde{\mathbf c}}}

\def\oneb{\ensuremath{{\mathbf 1}}\xspace}
\def\xs{\x^\star}
\def\A{{\mathbf A}}
\def\B{{\mathbf B}}
\def\C{{\mathbf C}}

\def\G{{\mathbf G}}
\def\H{{\mathbf H}}
\def\I{{\mathbf I}}
\def\M{{\mathbf M}}
\def\P{{\mathbf P}}
\def\U{{\mathbf U}}
\def\X{{\mathbf X}}
\def\tA{{\tilde{\mathbf A}}}
\def\tB{{\tilde{\mathbf B}}}
\def\tC{{\tilde{\mathbf C}}}

\def\Diag{\Lambda}

\newcommand{\Qc}{\mathcal{Q}}
\newcommand{\Rc}{\mathcal{R}}
\newcommand{\Qcs}{{\mathcal{Q}^\star}}

\def\R{\mathbb{R}}

\def\am{\arg\min}
\def\ama{\arg\max}

\def\vars_w{\sigma^2_n}
\def\vars{\sigma^2}
\def\ie{\textit{i.e.}, }
\def\eg{\textit{e.g.}, }
\def\etal{\textit{et al.} }
\def\spark{\mathrm{spark}}
\def\spa{\mathrm{span}}

\def\proj{\mathbf{P}_\Qc^\bot }

\def\AQp{\A_{\Qc'}}
\def\xQp{\x_{\Qc'}}
\def\ud{\bar{\delta}}
\def\ld{\underline{\delta}}

\usepackage{soul}

\newtheorem{defi}{Definition}

\newtheorem{proposition}{Proposition}
\newtheorem{lemma}{Lemma}
\newtheorem{theorem}{Theorem}

\newtheorem{example}{Example}

\begin{document}

\maketitle

%\remCS{Working assumptions: we need to assume that $spark(\A)\geq
%  j+1$ in order to define $\mu^{OLS}(j)$ and ensure that
%  $\lb(\ell,j)>0$. Can we reformulate Ths. 1 and 3 with the weaker
%  assumption $\A_{\Qc^\star}$ full rank or with no full rankness
%  assumption at all? It seems that $\mu < 1/(k-1)$ ensures that
%  $\tilde{\A}_{\Qc^\star\backslash\Qc}$ and
%  $\tilde{\B}_{\Qc^\star\backslash\Qc}$ are full rank (proof of Th.
%  2).}

\begin{abstract}
  We address the exact recovery of a $k$-sparse vector in the
  noiseless setting when some partial information on the support is
  available.  This partial information takes the form of either
    a subset of the true support or an approximate subset including
    wrong atoms as well.  We derive a new sufficient and worst-case
  necessary (in some sense) condition for the success of some
  procedures based on $\ell_p$-relaxation, Orthogonal Matching Pursuit
  (OMP) and Orthogonal Least Squares (OLS).
  Our result is based on the coherence $\mu$ of the dictionary and
  relaxes the well-known condition $\mu<1/(2k-1)$ ensuring the
  recovery of any $k$-sparse vector in the 
non-informed setup. It reads $\mu<1/(2k-g+b-1)$ when
    the informed support is composed of $g$ good atoms and $b$ wrong
    atoms. We emphasize that our condition
  is complementary to some restricted-isometry based conditions by showing that none
  of them implies the other.

  Because this mutual coherence condition is common to all
    procedures, we carry out a finer analysis based on the Null Space
    Property (NSP) and the Exact Recovery Condition (ERC). Connections
    are established regarding
  the characterization of $\ell_p$-relaxation procedures and OMP  in the informed setup. First, we emphasize that the truncated
  NSP enjoys an ordering property when $p$ is decreased. Second, the partial ERC for OMP
  (ERC-OMP) implies in turn the truncated NSP for the informed
  $\ell_1$ problem, and the truncated NSP for $p<1$.
 
% these generalized NSPs enjoy a nesting properties and can be related to the  partial Exact Recovery Condition (ERC) recently proposed by Soussen \etal. We then derive a sufficient condition based on the mutual coherence $\mu$ of the dictionary that guarantees the success of OMP, OLS, and any $\ell_p$-relaxation algorithm. In particular, our coherence-based result relaxes the well-known condition $\mu<1/(2k-1)$ to the case where the decoder has some partial knowledge of the support.
%  The latter condition is moreover shown to be worst-case necessary in some sense.
%  
  %We address the exact recovery of the support of a $k$-sparse vector with Orthogonal Matching Pursuit (OMP) and Orthogonal Least Squares (OLS) in a noiseless setting. We consider the scenario where OMP/OLS have selected good atoms during the first $l$ iterations ($l<k$) and derive a new sufficient and worst-case necessary condition for their success in $k$ steps. Our result is based on the coherence $\mu$ of the dictionary and relaxes Tropp's well-known condition $\mu<1/(2k-1)$ to the case where OMP/OLS have a \emph{partial} knowledge of the support.
\end{abstract}


\section{Introduction}

Sparse representations aim at describing a signal as the combination
of a few elementary signals (or atoms) taken from an overcomplete
dictionary $\A$. In particular, in a noiseless setting, one
wishes to find the vector with the smallest number of non-zero
elements, satisfying a set of linear constraints, that
is %. Specifically, one looks for the solution of
\begin{align}
\tag{$P_0$}
\min \| \x \|_0\quad \mbox{subject to $\A\x=\y$, } \label{eq:SRproblem}
\end{align}
where $\A\in\mathbb{R}^{m\times n}$, $\x\in\mathbb{R}^{n}$, $\y\in\mathbb{R}^{m}$. Unfortunately, problem \eqref{eq:SRproblem} is of combinatorial nature and, therefore, its resolution reveals  to be intractable in most practical settings \cite{Natarajan1995Sparse}. 

In order to address this issue, suboptimal (but tractable)
algorithms have been proposed in the literature. Among the most
popular procedures, let us mention: \emph{i)} the algorithms based on the $\ell_p$-relaxation of the $\ell_0$ pseudo-norm;
%
%
% to an $\ell_p$ norm ($p\in(0,1]$), \eg Basis Pursuit
%\cite{Chen_siam99}, FOCUSS \cite{Gorodnitsky_ieeetsp97}; 
%
\emph{ii)} the greedy algorithms, seen as sub-optimal discrete search algorithms to address ($P_0$).
On the one hand, the $\ell_p$ relaxation of \eqref{eq:SRproblem} can be expressed as
\begin{align}
\tag{$P_p$}
\min \| \x \|_p\quad \mbox{subject to $\A\x=\y$, } \label{eq:SRproblem_rel}
\end{align}
with $p\in(0, 1]$. Practical implementations of $(P_1)$, also named Basis Pursuit \cite{Chen_siam99}
can be done optimally using linear  programming algorithms, see \emph{e.g.,}~\cite{Elad2010};
sub-optimal procedures looking for a solution of $(P_p)$ with
$p\in(0,1)$ are for example derived in \cite{Gorodnitsky_ieeetsp97},
\cite{Foucart2009Sparsest}. On the other hand, greedy procedures
 build a sparse vector by gradually increasing the active subset
  starting from the empty set. At each iteration, a new atom is
  appended to the active subset. Standard greedy procedures include,
by increasing order of complexity,  Matching Pursuit
(MP)~\cite{Mallat_ieeetsp93}, Orthogonal Matching Pursuit
(OMP)~\cite{Pati_asilomar93}, Orthogonal Least Squares
(OLS)~\cite{Chen:1950fk, Natarajan1995Sparse} and variants thereof,
namely regularized OMP~\cite{Davenport2010Analysis}, weak
OMP~\cite{Foucart2013Stability}, stagewise
OMP~\cite{Donoho2012Sparse}, etc.

In this paper, we focus on a variation of the sparse representation problem in which the decoder has some information (possibly erroneous) about the support of the sparse vector. This new paradigm has recently been introduced independently in several contributions and finds  practical and analytical interests in many setups. 

In \cite{VonBorries2007Compressed, AminKhajehnejad2009Weighted, Vaswani2010ModifiedCS, Jacques2010Short,Friedlander2012Recovering,Bandeira2011Partially}, the authors focussed on the problem of recovering
a sequence of sparse vectors with a strong dependence on their supports. This type of settings occurs for example in video compression or dynamic magnetic resonance imaging where the support of the sought vectors commonly evolves slowly with time. 
 More specifically, this set of papers focusses on an $\ell_1$-relaxation of the following problem (or some slightly different variants thereof):
\begin{align}
\tag{$P_{0,\Qc}$}
\min_\x \| \x_{\bar{Q}} \|_0\quad \mbox{subject to $\A\x=\y$, } \label{eq:SRproblem_Q}
\end{align}
where $\Qc$ is an estimate of the sought support and $\x_{\bar{Q}}$ represents the vector made up of the elements of $\x$ whose index is not in $\Qc$.

More generally, the paradigm of sparse representation with side support information is of interest when some of the coefficients of the sparse decomposition can be easily identified a priori. For example, as mentioned in \cite{Jacques2010Short}, in wavelet image processing,  the coefficients weighting the scaling functions are likely to be non-zero and this information should  be (ideally) taken into account in any  processing. It also happens in many practical situations that some coefficients of the sparse decomposition (typically those with high amplitudes) can be identified by simple thresholding. This observation is the essence of the algorithm proposed  in \cite{Wang2010Sparse} where the authors look for a solution of \eqref{eq:SRproblem} by successively applying thresholding operations on the solution of $\ell_1$-relaxations of \eqref{eq:SRproblem_Q} to obtain a sequence of refined support estimates. 

A slightly different, but related, perspective was considered in 
\cite{Karahanoglu2012Theoretical} for OMP and in
\cite{Soussen2013Joint} for both
  OMP and OLS. In these papers, the authors derived guarantees of
success for OMP and OLS by assuming that atoms
belonging to some subset $\Qc$ have been selected during the first iterations. The
goal of such approaches is to provide a finer analysis of OMP/OLS at
intermediate iterations by noting that the standard uniform
recovery conditions ensuring the success of OMP/OLS from
the first iteration are rather pessimistic. It is quite
obvious that the conditions derived in these papers also apply to
situations where OMP/OLS are initialized with support $\Qc$ (rather
than with the empty support). In the sequel, we will refer to this
variant of OMP (resp. OLS) as OMP$_\Qc$ (resp. OLS$_\Qc$). Clearly,
OMP$_\Qc$/OLS$_\Qc$ can be understood as greedy procedures looking for
a solution of \eqref{eq:SRproblem_Q}.  

In this paper, we derive \emph{uniform} recovery conditions for OMP$_\Qc$/OLS$_\Qc$ and $\ell_p$-relaxed versions of \eqref{eq:SRproblem_Q} in the paradigm of partially-informed decoders. 
Our conditions are valid for $\y=\A\xs$ where $\xs$ is any $k$-sparse vector.  
Let us briefly summarize the related literature. 

First, generalizing the well-known ``Null-Space Property"  (NSP) derived in \cite{Gribonval2003Sparse}, the authors of \cite{Wang2010Sparse, Bandeira2011Partially, Chen2010Convergence} proposed a ``truncated" NSP, which is a sufficient and worst-case necessary condition for the success of 
\begin{align}
\tag{$P_{p,\Qc}$}
\min_\x \| \x_{\bar{Q}} \|_p\quad \mbox{subject to $\A\x=\y$, } \label{eq:SRproblemp_Q}
\end{align}
with $p\in[0,1]$. 
 Secondly, in \cite{Vaswani2010ModifiedCS,Jacques2010Short,Friedlander2012Recovering}, a series of sufficient conditions based on restricted isometry constants (RICs) were proposed to guarantee the success of \eqref{eq:SRproblemp_Q} (or some variants thereof). 
% In \cite{Vaswani2010ModifiedCS}, the authors showed that their RIC condition for exact reconstruction with ($P_{1,\Qc}$) is weaker than another RIC condition for ($P_1$) as long as $\Qc$ is a ``sufficiently accurate" estimate of the support of $\xs$. 
% This result was extended by Jacques to the cases of compressible signals and noisy observations in \cite{Jacques2010Short}. Finally, in \cite{Friedlander2012Recovering} Frielander \etal derived RIC conditions generalizing those derived in \cite{Candes_Romberg_Tao_2005,Candes2008Restricted} for the success of ($P_1$). A different point of view was considered in \cite{AminKhajehnejad2009Weighted} where the authors used a Grassman angle approach to characterize a class of signal which can be recovered by (a variant of) \eqref{eq:SRproblem_Q}.

Concerning OMP$_\Qc$/OLS$_\Qc$, the authors in \cite{Soussen2013Joint}
derived a partial ``Exact Recovery Condition" (ERC) extending Tropp's
ERC to the partially-informed paradigm considered in this
paper. The extended condition was shown to be sufficient but
also worst-case necessary for the success of OMP/OLS when some support
$\Qc$ has been selected at an intermediate iteration. 
 In \cite{Karahanoglu2012Theoretical}, the authors proposed a sufficient condition based on RICs and depending on the number of ``good" and ``bad" atoms selected in $\Qc$, that is the number of elements of $\Qc$ which are (resp. are not) in the support of $\xs$. 

In this paper, we derive a new simple recovery guarantee for OMP$_\Qc$, OLS$_\Qc$ and \eqref{eq:SRproblemp_Q} for $p\in[0,1]$. Our condition only depends on the mutual coherence of the dictionary and the number of good and bad atoms selected in the estimated support $\Qc$:
 \begin{align}
\mu<\frac{1}{2k-g+b-1} \label{eq:CBcondition_partial},
\end{align}
where $g$ (resp. $b$) denotes the number of ``good" (resp. ``bad")
atoms in $\Qc$. We show that \eqref{eq:CBcondition_partial} is
sufficient for the success of \eqref{eq:SRproblemp_Q} with
$p\in[0,1]$, and of OMP$_\Qc$ and OLS$_\Qc$. We emphasize moreover
that \eqref{eq:CBcondition_partial} is worst-case necessary in the
following sense: there exists a dictionary $\A$ with
$\mu=\frac{1}{2k-g+b-1}$, a combination $\y$ of $k$ columns of $\A$
and a support $\Qc$ containing $g$ good and $b$ bad atoms such that
neither \eqref{eq:SRproblemp_Q} nor OMP$_\Qc$/OLS$_\Qc$ can recover
$\xs$.  Our condition generalizes, within the
informed paradigm, the well-known condition $\mu<\frac{1}{2k-1}$
ensuring the success of Basis-Pursuit and OMP/OLS in the standard
setup, see \eg \cite{Tropp2004Greed,Gribonval2003Sparse,
  Gribonval2007Highly}. In particular, we see that if the informed
support $\Qc$ contains more than $50\%$ of good atoms,
\eqref{eq:CBcondition_partial} leads to a weaker condition than its
standard counterpart. On the other hand, although ensuring the success
of \eqref{eq:SRproblemp_Q} and OMP$_\Qc$/OLS$_\Qc$, condition
\eqref{eq:CBcondition_partial} does not allow for a discrimination of
the performance achievable by these algorithms.

In order to address this question, we analyze in this paper some
connections existing between the conditions previously proposed in the
literature. First, we show that the truncated NSP derived in
\cite{Wang2010Sparse, Bandeira2011Partially, Chen2010Convergence}
enjoys a nesting property, namely: if the truncated NSP is satisfied
for some $p\in[0,1]$, then it is also verified for any other $q\in
[0,p]$. From a worst-case point of view, this result tends to show
that the resolution of \eqref{eq:SRproblemp_Q} with $p\in[0,1)$ is
more favorable than $\ell_1$-based approaches\footnote{We note however
  that, unlike the convex $\ell_1$ problem, reaching the global
  minimum of $\ell_p$ problems is not guaranteed in practice.}. In particular, as a
corrolary of this result, we have that all uniform conditions
previously proposed for $(P_{1,\Qc})$
also guarantee the success of \eqref{eq:SRproblemp_Q} with
$p\in[0,1)$.  Second, we establish that the partial ERC derived in
\cite{Soussen2013Joint} for OMP$_\Qc$ is also a sufficient condition
for the success of ($P_{1,\Qc}$). This generalizes the result derived
by Tropp in the standard (non-informed) setup \cite{Tropp2004Greed} to
the partially-informed context considered in this paper. On the other
hand, we emphasize that, unlike in the standard setup, such a
connection does not hold between ($P_{1,\Qc}$) and OLS$_\Qc$. 
 
Finally, we also study the connection between the proposed
coherence-based condition \eqref{eq:CBcondition_partial} and some
RIC-based conditions previously proposed in the context of 
orthogonal greedy algorithms. First, we illustrate the
complementarity of \eqref{eq:CBcondition_partial} with the RIC
guarantees proposed in \cite{Karahanoglu2012Theoretical} for OMP. We emphasize
that no condition implies the other one. Secondly, we show that the
RIC condition proposed in \cite{Karahanoglu2012Theoretical} for the
success of OMP$_\Qc$ also enjoys a form of quasi-tightness for
both OMP$_\Qc$ and OLS$_\Qc$.
 
% Third, we will generalize the coherence-based condition \eqref{eq:CBcondition} to the case where some partial information on the support of $\xs$ is available. Our coherence-based recovery condition writes
% \begin{align}
%\mu<\frac{1}{2k-g+b-1} \label{eq:CBcondition_partial},
%\end{align}
%where $g$ (resp. $b$) denotes the number of  ``good" (resp. ``bad") atoms in $\Qc$. We will show that \eqref{eq:CBcondition_partial} is a sufficient  recovery condition for \eqref{eq:SRproblemp_Q} with $p\in[0,1]$,  and for OMP/OLS when initialized with support $\Qc$. Some form of worst-case necessity of \eqref{eq:CBcondition_partial} will also be emphasized. Finally, the complementarity of \eqref{eq:CBcondition_partial} with a RIC-based condition derived in \cite{Karahanoglu2012Theoretical} will be pointed out. 
 
The rest of this paper is organized as follows. In section \ref{sec:notations}, we set the notations which will be used throughout the paper. 
  In section \ref{sec:Oxxdef}, we review the main expressions defining
  the recursions of OMP/OLS and briefly discuss their application to
  the informed problem \eqref{eq:SRproblem_Q}. Our contributions and
  their positioning within the current
  state-of-the-art are discussed in section
  \ref{sec:mainresults}. Finally, the remaining sections and
  appendices are dedicated to the proofs of our results.  
 
\section{Notations}\label{sec:notations}

The following notations will be used in this paper.
$\stdscal{\,.\,,\,.\,}$ refers to the inner product between vectors,
$\|\,.\,\|$ and $\|\,.\,\|_1$ stand for the Euclidean and the $\ell_1$
norms, respectively. $\|\,.\,\|_0$ will denote the $\ell_0$ pseudo norm which counts the number of non-zero elements in its argument. With a slight abuse of notation and for the sake of conciseness in some of our statements, we will also assume that $\|\,.\,\|_0^0\triangleq\|\,.\,\|_0$. 
 $.^\dag$ denotes the pseudo-inverse of a matrix.  For a full rank and undercomplete matrix $\X$, we have
$\X^\dag=(\X^T\X)^{-1}\X^T$ where $.^T$ stands for the matrix
transposition. When $\X$ is overcomplete, $\spark (\X)$ denotes the
minimum number of columns from $\X$ that are linearly
dependent~\cite{Donoho2003Optimally}. $\mathbf{1}_{m}$ (resp
$\mathbf{0}_{m}$) denotes the all-one (resp. all-zero) vector of
dimension $m\times 1$. $\I_m$ is the $m\times m$ identity matrix. The letter $\Qc$ denotes some subset of the column
indices and $\bar{\Qc}$ is the complementary set of $\Qc$. $\X_{\Qc}$ is the submatrix of $\X$ gathering the columns
indexed by $\Qc$. For vectors, $\x_\Qc$ denotes the subvector
  of $\x$ indexed by $\Qc$.  We will denote the cardinality of $\Qc$
as $\vert \Qc \vert$. We use the same notation to denote the absolute
value of a scalar quantity. For $\X_\Qc\in\R^{m\times k}$, $\mathbf{P}_\Qc=\X_\Qc\X_\Qc^\dag$ and $\proj=\I_m-\mathbf{P}_\Qc$
denote the orthogonal projection operators onto $\spansub{\X_\Qc}$ and
$\spansub{\X_\Qc}^{\perp}$, where $\spansub{\X}$ stands for the column
span of $\X$, $\spansub{\X}^{\perp}$ is the orthogonal complement of
$\spansub{\X}$. $\r^{\Qc}=\P^\perp_{\Qc} \y=\y-\P_{\Qc}\y$ denotes the data residual
  induced
by the orthogonal projection of $\y$ onto $\spansub{\X_\Qc}$. Finally, we will use the notation $\ker(\X)\triangleq\spansub{\X^T}^\perp$ to denote the null space of $\X$; $\ker_0(\X)$ is the null-space of $\X$ minus the all-zero vector.

\section{OMP and OLS} \label{sec:Oxxdef}

In this section, we recall the selection rules defining OMP and OLS, and discuss their application to the support-informed problem \eqref{eq:SRproblem_Q}. Throughout the paper, we will use the common acronym Oxx in statements that apply to both OMP and OLS.

First note that any vector $\x$ satisfying the constraint in
\eqref{eq:SRproblem} must have a support, say $\tilde{\Qc}$, such that
$\P^\perp_{\tilde{\Qc}} \y =\mathbf{0}_m$ since $\y$ must belong to
$\spansub{{\A}_{\tilde{\Qc}}}$. Hence, problem \eqref{eq:SRproblem} can
equivalently be rephrased as
\begin{align}
\min_{\tilde{\Qc}} \vert \tilde{\Qc} \vert \quad \mbox{subject to $\P^\perp_{\tilde{\Qc}} \y =\mathbf{0}_m$. } \label{eq:SRproblem2}
\end{align}
Oxx can be understood as an iterative procedure searching for a solution of \eqref{eq:SRproblem2} by generating a sequence of support estimates $\{\Qc^{(\ell)}\}$  as
%adding one new element to the current support as
\begin{align}
\Qc^{(\ell+1)} =  \Qc^{(\ell)} \cup \{j\},
\end{align}
where
\begin{align}\label{eq:atomselection1}
j \in \left\{
\begin{array}{ll}
\ama_i  \vert \stdscal{\a_i,\r^{\Qc^{(\ell)}}} \vert  & \textrm{for OMP}\\
\am_i \| \r^{\Qc^{(\ell)} \cup \{i\}}\|& \textrm{for OLS,}
\end{array}\right. 
\end{align}
$\r^{\Qc^{(\ell)}}\triangleq\P^\perp_{\Qc^\ell} \y$ is the current data residual and $\a_i$ is the $i$th column of $\A$. More specifically, Oxx adds
one new atom to the support at each iteration: OLS selects the atom
minimizing the norm of the new residual $\r^{\Qc^{(\ell)}\cup\{i\}}$
whereas OMP picks the atom maximizing the correlation with the current
residual.

Oxx is commonly initialized with the empty set, \ie
$\Qc^{(0)}=\emptyset$. However, when some initial estimate
($\Qc$) of the support of $\xs$ is
available, nothing prevents us from initializing Oxx
with $\Qc^{(0)}=\Qc$.  We will refer to this variant of Oxx as
Oxx$_\Qc$. Note that at the first iteration of Oxx$_\Qc$, the
  redisual is initialized by $\r^{\Qc}\triangleq \P^\perp_{\Qc}\y$,
  \ie the data $\y$ are being projected onto $\spansub{\A_{\Qc}}^\bot$.
  In other words, Oxx$_\Qc$ behaves similarly with $\y$ or
  $\P^\perp_{\Qc}\y$ as input vector.
 %Obviously, Oxx$_\Qc$ reduces to the standard implementation of Oxx as soon as $\Qc=\emptyset$. Moreover,
 
On the one hand, Oxx$_\Qc$ can readily be seen as a
greedy procedure looking for a solution of \eqref{eq:SRproblem_Q}. On
the other hand, the behavior of Oxx$_\Qc$ can be
understood from a different perspective, namely the analysis of Oxx at
an intermediate iteration to address ($P_{0,\emptyset}$).
Indeed, let us assume that Oxx has selected atoms in $\Qc$ during the
first $\vert \Qc\vert$ iterations. Then, the next step of Oxx will be
identical to the first iteration of Oxx$_\Qc$.  Although we will
mainly stick to the former vision
hereafter%(\ie Oxx$_\Qc$ as a greedy implementation of \eqref{eq:SRproblem_Q})
, the results that will be derived in the paper can be interpreted
from these two perspectives.

In the sequel, we will often use a slightly different, equivalent, formulation of \eqref{eq:atomselection1} based on orthogonal projections. Let us define
\begin{align}
\ta_i & \triangleq \P_{\Qc^{(\ell)}}^\perp \a_i, \\
\tb_i & \triangleq \left\{
\begin{array}{ll}
{\ta_i}/{\| \ta_i \|}& \mbox{if $\ta_i\neq\mathbf{0}_m$}\\
\mathbf{0}_m & \mbox{otherwise.}
\end{array}\right. \label{eq:defbi}
\end{align}
$\ta_i$ denotes the projection of $\a_i$ onto $\spansub{\A_{\Qc^{(\ell)}}}^\bot$ whereas $\tb_i$ is a normalized version of $\ta_i$. 
 For simplicity, we dropped the dependence of $\ta_i$ and $\tb_i$ on $\Qc^{(\ell)}$ in our notations. However, when there is a risk of confusion, we will use $\ta_i^{\Qc^{(\ell)}}$ (resp. $\tb_i^{\Qc^{(\ell)}}$) instead of $\ta_i$ (resp. $\tb_i$). With these notations, \eqref{eq:atomselection1} can be re-expressed as 
\begin{align}\label{eq:atomselection2}
j \in \left\{
\begin{array}{ll}
\ama_i  \vert \stdscal{\ta_i,\r^{\Qc^{(\ell)}}} \vert  & \textrm{for OMP}\\
\ama_i \vert \stdscal{\tb_i,\r^{\Qc^{(\ell)}}} \vert & \textrm{for OLS}.
\end{array}\right.
\end{align}
The equivalence between \eqref{eq:atomselection1} and
\eqref{eq:atomselection2} is straightforward for OMP by noticing that
$\r^{\Qc^{(\ell)}} \in \spansub{\A_{\Qc^{(\ell)}}}^\bot$. We refer the reader to
\cite{RebolloNeira2002Optimized} for a detailed calculation for OLS.

Moreover, we will define the  unifying notation:
\begin{align}
\tc_i \triangleq
  \left\{
    \begin{array}{ll}
      \ta_i & \textrm{for OMP},\\
      \tb_i & \textrm{for OLS},
    \end{array}
  \right.
  % \label{eq:ci}
\end{align}
and use the notations $\tA$, $\tB$ and $\tC$ to refer to the matrices whose columns are made up of the $\ta_i$'s, $\tb_i$'s and $\tc_i$'s, respectively.

\section{Context and Main Results}\label{sec:mainresults}

Let us assume that $\y$ is a linear combination of $k$ columns of $\A$, that is 
%\begin{align}
%\y = \A_\Qcs \x_\Qcs \qquad \mbox{with $\vert \Qcs \vert=k$, $\,x_i\neq 0\  \forall i\in \Qcs$}.
%\end{align}
%
\begin{align}
\y = \A \xs \qquad \mbox{with $\,x^\star_i\neq 0\Leftrightarrow  \ i\in \Qcs$,  $\vert \Qcs \vert=k$}.\label{eq:defy}
\end{align}
 In this section, we review some standard conditions ensuring the correct reconstruction of $\xs$ (with and without partial information on the support) and recast our contributions within these existing results. We will use the following conventions: 
the atoms $\a_i$ in $\Qcs$ will be referred to as ``good'' atoms whereas atoms not in $\Qcs$ will be dubbed ``bad" atoms. If an initial estimate of the support  $\Qcs$ is available, say $\Qc$, we will denote by $g\triangleq\vert \Qc\cap\Qcs \vert$ the number of good atoms in $\Qc$ and by $b\triangleq\vert \bar{\Qc}^\star\cap\Qc \vert$ the number of bad atoms.
% in $\Qc$. 
We will always  implicitly assume that $g<k$ since otherwise the informed problem $(P_{\Qc,0})$ becomes trivial. 
 Finally, we will suppose that the columns of $\A$ are normalized throughout the paper.

Our contributions will be both at the level of Oxx$_\Qc$ and $(P_{p,\Qc})$. In the next subsection we will focus on the conditions pertaining to Oxx and Oxx$_\Qc$ whereas in subsection \ref{ssec:ContribPpQ}, we will describe the guarantees associated to the success of \eqref{eq:SRproblem_rel} and \eqref{eq:SRproblemp_Q}. 
%In the next two subsections, we distinguish between the  contributions pertaining to these two different algorithmic approaches and state our main results.
 Let us mention that our contributions are uniform conditions derived within the context of worst-case analyses. Hence, hereafter, we will essentially limit our discussion to the contributions in this line of thought.  

Before proceeding, we recall the standard definitions of the restricted isometry constant (RIC) and mutual coherence that will be used in our discussion:
% We recall the definitions of these two quantities hereafter:
\begin{defi}%[$k$-th order restricted isometry constant] 
The $k$-th order restricted isometry constant of $\A$ is the smallest non-negative value $\delta_k$ such that the following inequalities
\begin{align}
(1-\delta_k) \| \x\|^2 \leq \| \A\x\|^2 \leq (1+\delta_k)\| \x\|^2
\end{align}
are verified for any $k$-sparse vector $\x$.
\end{defi} 
\begin{defi}%[Mutual coherence] 
\label{def:mu} The mutual coherence $\mu$ of a dictionary $\A$ is  defined as
\begin{align}
\mu = \max_{i \neq j} \vert \stdscal{\a_i,\a_j}\vert.
\end{align}
\end{defi}

\subsection{Results and state-of-the-art conditions for Oxx and Oxx$_\Qc$}\label{ssec:ContribOxx}

OMP has been widely studied in the recent years, including worst
case~\cite{Tropp2004Greed,Davenport2010Analysis} and probabilistic
analyses~\cite{Tropp2007Signal}. The existing exact recovery analyses
of OMP were also adapted to several extensions of OMP, namely
regularized OMP~\cite{Davenport2010Analysis}, weak
OMP~\cite{Foucart2013Stability}, and stagewise
OMP~\cite{Donoho2012Sparse}.  Although OLS has been known in the
literature for a few decades (often under different names
\cite{Blumensath2007Difference}), exact recovery analyses of OLS
remain rare for two reasons. First, OLS is significantly more time
consuming than OMP, therefore discouraging the choice of OLS for
``real-time'' applications, like in compressive sensing. Secondly, the
selection rule of OLS is more complex, as the projected atoms are
normalized. This makes the OLS analysis more tricky. When the
dictionary atoms are close to orthogonal, OLS and OMP have a similar
behavior, as emphasized in~\cite{Foucart2013Stability}. On the
contrary, for correlated dictionary (\eg in ill-conditioned inverse
problems), their behavior significantly differ and OLS may be a better
choice~\cite{Soussen2013Joint}.  The above arguments motivate our
analysis of both OMP and OLS, interpreted as sub-optimal
  algorithms to address ($P_0$).
%  although in the present paper, our low mutual coherence assumptions
%  imply that the correlation between atoms is weak, therefore we do
%  not exhibit any difference of behavior between OMP and OLS. 

Let us first rigorously define the notion of ``success" that will be used for Oxx$_\Qc$ throughout the paper:
\begin{defi}[Successful recovery] \label{def:success}
Oxx$_\Qc$ with $\y$ defined in \eqref{eq:defy} as input succeeds if and only if it selects atoms in $\Qcs\backslash\Qc$ during the  first $k-g$ iterations. 
\end{defi}
In particular, this definition implies that Oxx$_\Qc$ exactly reconstructs $\xs$ after $k-g$ iterations, as long as $\A_{\Qcs\cup\Qc}$ is full rank. 

We will moreover assume that, in special cases where the Oxx$_\Qc$ selection rule yields multiple solutions including a wrong atom, that is
 \begin{align}\label{eq:equalitySR}
\max_{i\in\Qcs} \vert \stdscal{\tc_i,\r^{\Qc^{(\ell)}}}  \vert = \max_{i\notin\Qcs} \vert \stdscal{\tc_i,\r^{\Qc^{(\ell)}}}  \vert,
\end{align}
Oxx$_\Qc$ systematically takes a bad decision. Hence, situation~\eqref{eq:equalitySR} always leads to a recovery failure. 

Remember that Oxx$_\Qc$ reduces to the standard implementation of Oxx as soon as $\Qc=\emptyset$. In this case, Definition \ref{def:success} matches the classical ``$k$-step" analysis encountered in many contributions of the literature. Let us however mention that the notion of successful recovery may be defined in a weaker sense:
Plumbley~\cite[Corollary~4]{Plumbley2007Polar} first pointed
out that there exist problems for which ``delayed recovery'' occurs
after more than $k$ steps. Specifically, Oxx can select some wrong
atoms during the first $k$ iterations but ends up with a larger
support including $\Qc^\star$ with a number of iterations
  slightly greater than $k$. In the noise-free setting (for
  $\y\in\spansub{\A_{\Qcs}}$), all atoms not belonging to
$\Qc^\star$ are then weighted by 0 in the solution vector
(under full rank assumptions).
Recently, a delayed recovery analysis of OMP using restricted-isometry
constants was proposed in~\cite{Zhang2011Sparse} and then extended to
the weak OMP algorithm (including OLS) in~\cite{Foucart2013Stability}.

To some extent, the definition of success considered in this paper
also partially covers the setup of delayed recovery. Indeed, keeping
in mind that Oxx$_\Qc$ can be understood as a particular instance of
Oxx in which atoms in $\Qc$ have been selected during the first $g+b$
iterations, any condition ensuring the success of Oxx$_\Qc$ in the
sense of Definition \ref{def:success} also guarantees the success of
Oxx in $k+b$ iterations as long as atoms in $\Qc$ are selected during
the first $g+b$ iterations. Conditions under which $g$ good and $b$ bad atoms are selected during the first iterations are however not discussed in the rest of the paper. 
%\remCH{Rediscuter Avec Charles}However, it is worth mentioning that, although shedding some light on the understanding of Oxx's delayed recovery, our coherence-based result stated herefater (see Theorem \ref{th:mainth}) does not weaken Tropp's condition for $k$-step recovery in the non-informed setup, that is from an initial empty support (see Theorem \ref{th:SCmu}).
%\remCS{provided that $g\geq b$?} \remCH{Ici je voulais dire qu'au point de vue du succès de Oxx à partir de la première itération, notre résultat ne fait pas mieux que celui de Tropp. En effet, notre condition devient plus faible que celle de Tropp a partir du moment où Oxx a sélectionné plus de bons atomes que de mauvais. Toutefois nous ne nous préoccupons pas de la caractérisation de ce genre de situations et dans le cas où $g=b=0$ on retrouve exactement le même résultat que Tropp. }

 Regarding $k$-step analyses, the first thoughtful theoretical study of OMP is due to Tropp, see
\cite[Th. 3.1 and Th. 3.10]{Tropp2004Greed}. Tropp provided a
sufficient and worst-case necessary condition for the exact recovery
of any sparse vector with \emph{a given} support $\Qcs$. The
derivation of a similar condition for OLS is more recent and is due to
Soussen \etal in \cite{Soussen2013Joint}. In the latter paper, the
authors carried out a narrow analysis of both OMP and OLS at
any intermediate iteration of the algorithms. Their recovery
  conditions depend not only on $\Qcs$ but also on the support $\Qc^{(\ell)}$ estimated by Oxx at a given iteration $\ell$.  Recasting this analysis  within the framework of sparse recovery with partial support information, $\Qc^{(\ell)}$ plays the role of the estimated support $\Qc$,   and the main result in~\cite{Soussen2013Joint} can be rewritten as:

\begin{theorem}[Soussen \etal 's Partial ERC {\cite[Th. 3]{Soussen2013Joint}}] \label{th:SoussenERC}Assume that $\A_{\Qcs\cup\Qc}$ is full rank with $\vert \Qcs\vert=k$, $\vert \Qcs\cap\Qc\vert=g<k$, and $\vert \bar{\Qc}^\star\cap\Qc\vert=b$.
 If %Oxx with $\y\in\spansub{\A_{\Qcs}}$ as input selects atoms in $\Qc$ during the first $l$ iterations, and
 \begin{align}
\max_{i \notin \Qcs} \| \tC_{\Qcs \backslash \Qc}^\dag \tc_i \|_1 < 1, \label{eq:SoussenERC}
\end{align}
then
for any $\y\in\spansub{\A_{\Qcs}}$,  Oxx$_\Qc$ only selects atoms in $\Qcs \backslash \Qc$ during the first 
 $k-g$ iterations. Conversely, if
\eqref{eq:SoussenERC} does not hold, there exists $\y
\in\spansub{\A_\Qcs}$  for which Oxx$_\Qc$ selects a bad atom $j\notin\Qcs$ at the first iteration.
\end{theorem}
%
%We note that \eqref{eq:SoussenERC}, on its own, does not constitute a worst-case necessary condition for OMP if $ \Qc \neq \emptyset$. More specifically, as shown in \cite{Soussen2013Joint}, some additional ``reachability" hypotheses are required for \eqref{eq:SoussenERC} to be a worst-case necessary condition for OMP. 
% We refer the reader to \cite{Soussen2013Joint} for a precise definition of the notion of  ``reachability". %Theorem \ref{th:SoussenERC} provides sufficient and worst-case necessary condition for Oxx at any iteration of the algorithm.  
%
The proof of Theorem \ref{th:SoussenERC} is a straightforward adaptation
of~\cite[Theorem 3]{Soussen2013Joint}. For conciseness reasons, we
decide to skip it. Let us just mention that the original formulation
of~\cite[Theorem 3]{Soussen2013Joint} involves a vector
$\y\in\spa(\A_{\Qcs\cup\Qc})$. Because any vector
$\y\in\spa(\A_{\Qcs\cup\Qc})$
can be uniquely decomposed as $\y=\y_1+\y_2$ with
$\y_1\in\spa(\A_{\Qcs})$, 
$\y_2\in\spa(\A_{\bar{\Qc}^\star\cap\Qc})$ under full rank conditions,
and because Oxx$_\Qc$ has the same behavior with $\y$ and $\y_1$ as inputs
(the component $\y_2$ indexed by $\Qc$ is not taken into account),
both sufficient and necessary parts in Theorem \ref{th:SoussenERC} involve data vectors
$\y\in\spa(\A_{\Qcs})$.

Interestingly, when $\Qc=\emptyset$, Theorem \ref{th:SoussenERC} reduces to Tropp's ERC \cite{Tropp2004Greed}:
\begin{align}
\max_{i\notin \Qcs} \| \A_{\Qcs}^\dag \a_i \|_1 < 1, \label{eq:TroppERC}
\end{align}
which constitutes a sufficient and worst-case necessary condition for Oxx when no support information is available (or, equivalently, at the very first iteration of the algorithm). 
%In the general case $\Qc\neq \emptyset$, \eqref{eq:SoussenERC} is a weaker condition than \eqref{eq:TroppERC} which ensures an exact support recovery during the last iterations of Oxx. 

%One drawback of Tropp's and Soussen \etal 's ERCs stands in their cumbersome evaluation. 
%Indeed, evaluating \eqref{eq:SoussenERC}-\eqref{eq:TroppERC} requires to carry out a pseudo-inverse (and a projection for \eqref{eq:SoussenERC}) operation. 
A tight condition for the recovery of any
 $k$-sparse vector from any support estimate $\Qc$ such that $\vert \Qcs\cap\Qc\vert=g$, $\vert \bar{\Qc}^\star\cap\Qc\vert=b$ can therefore be expressed as
\begin{align}
\theta_{\mathrm{Oxx}}(k,g,b)<1,
\end{align}
where
\begin{align}
\theta_{\mathrm{Oxx}}(k,g,b)\triangleq\max_{\vert \Qcs\vert=k}\max_{\substack{\vert \bar{\Qc}^\star\cap\Qc\vert=b \\ \vert \Qcs\cap\Qc\vert=g}}\Bigl\{ \max_{i \notin \Qcs} \| \tC_{\Qcs \backslash \Qc}^\dag \tc_i \|_1\Bigr\}.\label{eq:ERCunif}
\end{align}
Unfortunately, the main drawback of \eqref{eq:ERCunif} stands in its cumbersome (combinatorial) evaluation.  
 In order to circumvent this issue, stronger conditions, but easier
to evaluate, have been proposed in the literature. We can mainly
distinguish between two types of ``practical" guarantees: the
conditions based on restricted-isometry constants and those
based on the mutual coherence of the dictionary.

The
contributions~\cite{Davenport2010Analysis,Huang2011Recovery,Liu2012Orthogonal,DBLP:journals/corr/abs-1102-4311,Mo2012Remark,Wang2012Recovery}
provide RIC-based sufficient conditions for the exact recovery of the
support $\Qcs$ in $k$ steps by OMP. The most recent and tightest results are
due to Maleh \cite{DBLP:journals/corr/abs-1102-4311}, Mo\&Shen
\cite{Mo2012Remark} and Wang\&Shim \cite{Wang2012Recovery}. The authors proved that OMP succeeds in $k$ steps
if 
\begin{align}
\delta_{k+1}<\frac{1}{\sqrt{k}+1}. \label{eq:RIP_k+1}
\end{align}
%where $\delta_{k+1}$ is the $(k+1)$-RIC of $\A$. 
In \cite[Th. 3.2]{Mo2012Remark} and \cite[Example 1]{Wang2012Recovery}, the authors
showed moreover that this condition is almost tight, \ie there exists
a dictionary $\A$ with $\delta_{k+1}=\frac{1}{\sqrt{k}}$ and a
$k$-term representation $\y$ for which OMP 
%of $k$ columns of $\A$, such that OMP with $\y$ as input 
selects a wrong atom at the first iteration. Let us mention that, by
virtue of Theorem \ref{th:SoussenERC}, these results remain valid for
OLS. Indeed, when $\Qc=\emptyset$,
  \eqref{eq:TroppERC} is a worst-case necessary condition of exact
  recovery for both OMP and OLS. Since~\eqref{eq:RIP_k+1} is a uniform
  sufficient condition for OMP, \eqref{eq:RIP_k+1} implies
  \eqref{eq:TroppERC}. 
 Very recently, Karahanoglu and Erdogan
\cite{Karahanoglu2012Theoretical} showed that
the condition
\begin{align}
\delta_{k+b+1}<\frac{1}{\sqrt{k-g}+1}\label{eq:RICCOMPQ}
\end{align}
is sufficient for the success of OMP$_\Qc$ when some support
information is available at the decoder. Similar conditions are still
not available for OLS$_\Qc$ and remain an open problem in the
literature.

In this paper, we emphasize that the RIC-based condition
\eqref{eq:RICCOMPQ}  also enjoys a type of worst-case
necessity. In particular, the following result shows that
\eqref{eq:RICCOMPQ} is almost tight for the success of OMP$_\Qc$ in
the following sense:
 \begin{lemma}[Quasi worst-case necessity of \eqref{eq:RICCOMPQ} for Oxx]\label{lem:quasi-tighnessRIC}
There exists a dictionary $\A$, a $k$-term representation $\y$ and  a set $\Qc$ with $\vert\Qcs\cap\Qc\vert=g$ and $\vert\bar{\Qc}^\star\cap\Qc\vert=b$, 
such  that: \emph{(i)} $\delta_{k+b+1}=\frac{1}{\sqrt{k-g}}$;
  \emph{(ii)} Oxx$_\Qc$ with $\y$ as input selects a bad atom at the first iteration.
  \end{lemma}
The proof of this lemma is reported to section \ref{sec:quasitighnessRIC}. 
Let us mention that the result stated in Lemma \ref{lem:quasi-tighnessRIC} is valid for both OMP and OLS. Hence, although the bound \eqref{lem:quasi-tighnessRIC} has not been proved to be a sufficient condition for the success of OLS$_\Qc$, this example shows that one cannot expect to  achieve much better guarantees in terms of RICs for OLS. 

%We will show in section \ref{ssec:Relationship} that \eqref{eq:RICCOMPQ} is also almost tight (in some  sense) for the success of OMP$_\Qc$ (see Theorem \ref{th:RICOMPQtight}). %: there exists a dictionary $\A$ with $\delta_{k+b+1}=\frac{1}{\sqrt{k-g}}$, a $k$-term representation $\y$ and a support $\Qc$ with $\vert\Qc\backslash\Qcs\vert=b$ and $\vert\Qc\cap\Qcs\vert=g$, such that OMP$_\Qc$ selects a bad atom at the first iteration. \emph{Situer la contribution dans le papier.}

Regarding uniform conditions based on the mutual coherence of the dictionary, Tropp showed in \cite[Cor. 3.6]{Tropp2004Greed} that 
\begin{align}
  \mu < \frac{1}{2k-1} \label{eq:CBcondition}
\end{align}
is sufficient for the success of OMP in $k$ steps. As a matter of
fact, \eqref{eq:CBcondition} therefore ensures that
\eqref{eq:TroppERC} is satisfied and thus also guarantees the success
of OLS (Theorem \ref{th:SoussenERC} with $\Qc=\emptyset$).  Moreover,
Cai\&Wang recently showed in \cite[Th. 3.1]{Cai2010Stable} that
\eqref{eq:CBcondition} is also worst-case necessary in the following
sense: there exists (at least) one $k$-sparse vector $\xs$ and one
dictionary $\A$ with $\mu=\frac{1}{2k-1}$ such that Oxx\footnote{and
  actually, any sparse representation algorithm. } cannot recover
$\xs$ from
$\y=\A\xs$. %Finally, the work by Soussen's \etal (see Theorem \ref{th:SoussenERC}) implies that \eqref{eq:CBcondition} is also sufficient and worst-case necessary for OLS.
These results are summarized in the following theorem: 
\begin{theorem}%[$\mu$-based uniform condition for Oxx {\cite[Cor. 3.6]{Tropp2004Greed}}, {\cite[Th. 3.1]{Cai2010Stable}}]\label{th:SCmu}
\textnormal{\textbf{[$\mu$-based uniform condition for Oxx {\cite[Cor. 3.6]{Tropp2004Greed}}, {\cite[Th. 3.1]{Cai2010Stable}}]\label{th:SCmu}}}
If \eqref{eq:CBcondition} is satisfied, then Oxx succeeds in
recovering any $k$-term representation.
%support $\Qcs$ with $\vert \Qcs \vert=k$ for any $\x_\Qcs\in
%\mathbb{R}^{k}$. 
Conversely, there exists an instance of dictionary $\A$ and 
a $k$-term representation for which:
%support $\Qcs$, with $\vert \Qcs \vert=k$, such that: 
\emph{(i)} $\mu=\frac{1}{2k-1}$; \emph{(ii)} Oxx selects a wrong atom
%$j \notin \Qcs$ 
at the first iteration.
\end{theorem}

%{\color{blue} Ubiquity of \eqref{eq:CBcondition} in the guarantees of performance of sparse-representation algorithms? (noisy cases, Basis Pursuit).} 

In this paper, we extend the work by Soussen \etal and provide a
coherence-based sufficient and worst-case necessary condition for the
success of Oxx$_\Qc$. Our  result
generalizes Theorem \ref{th:SCmu} as follows:
%in $\Qcs$ 

\begin{theorem}[$\mu$-based uniform condition for Oxx$_\Qc$]\label{th:mainth}
  Consider a $k$-term representation $\y
    =\A\xs$ and a subset $\Qc$ such that  $\vert\Qcs\cap\Qc\vert=g$ and $\vert\bar{\Qc}^\star\cap\Qc\vert=b$. If $\mu<\frac{1}{2k-g+b-1}$ holds, 
%    and 
%%, with $\vert \Qc\vert=l$, $\vert \Qcs\vert=k$. 
%\begin{align}
%\mu < \frac{1}{2k-g+b-1}. \label{eq:mainBound}
%\end{align}
then  Oxx$_\Qc$ recovers $\xs$ in $k-g$ iterations. 
Conversely, there exists a dictionary $\A$ and a $k$-term
  representation $\y$ such that: \emph{(i)} $\mu=\frac{1}{2k-g+b-1}$;
  \emph{(ii)} Oxx$_\Qc$ with $\y$ as input selects a bad atom at the first iteration.
% for at least one \addCS{vector
%   $\x_\Qcs\in\mathbb{R}^{k}$.

% supports $\Qc$, $\Qcs$, with $\vert \Qc \vert=l$, $\vert \Qcs
% \vert=k$, $\Qc \subset \Qcs$, such that: \emph{(i)}
% $\mu=\frac{1}{2k-l-1}$; \emph{(ii)} Oxx selects atoms in $\Qc$
% during the first $l$ iterations and then a wrong atom $j \notin
% \Qcs$ at the $(l+1)$th iteration for at least one \addCS{vector
%   $\x_\Qcs\in\mathbb{R}^{k}$.
\end{theorem}

The proof of this theorem is reported to sections \ref{sec:SCOMP},
\ref{sec:SCOLS} and \ref{sec:NC}. More specifically, we show in
section \ref{sec:SCOMP} (resp. section \ref{sec:SCOLS}) that
\eqref{eq:CBcondition_partial} is sufficient for the success of OMP$_\Qc$ (resp. OLS$_\Qc$)
in $k-g$ iterations. The proof of this sufficient
condition significantly differs for OMP$_\Qc$ and OLS$_\Qc$. The result is shown
for OMP$_\Qc$ by deriving an upper bound on Soussen~\etal's 
ERC-OMP condition \eqref{eq:SoussenERC} as
a function of the restricted isometry bounds of the projected
dictionary $\tA$. As for OLS$_\Qc$, the proof is based on a connection between
Soussen~\etal's ERC-OLS condition \eqref{eq:SoussenERC} and the mutual coherence of the normalized
projected dictionary $\tB$. Finally, in section \ref{sec:NC} we prove
that \eqref{eq:CBcondition_partial} is worst-case necessary for Oxx in the sense
specified in Theorem \ref{th:mainth}. This proof is common to both OMP$_\Qc$
and OLS$_\Qc$. If $b=0$, we also prove a slightly stronger result by showing that the subset $\Qc$ appearing in the converse part of Theorem \ref{th:mainth} can indeed be ``reached" by Oxx, initialized with the empty support.
%\remCS{cette phrase n'est pas claire: quels sont ces interets? Enlever cette fin de phrase pour simplifier?}
%which have some interests in the analysis of Oxx  at intermediate iterations.  We show that the subset $\Qc$ appearing in the converse part of Theorem \ref{th:mainth} is indeed ``reach\addCS{ed}\suppCS{able}" by Oxx. 
 More formally, the following result holds:
 \begin{lemma}[$\mu$-based partial uniform condition for Oxx]\label{lem:Oxxwc2}
There exists a dictionary $\A$, a $k$-term representation $\y$ and  a set $\Qc\subset\Qcs$ with $\vert\Qc\vert=g$, 
such  that: \emph{(i)} $\mu=\frac{1}{2k-g-1}$;
  \emph{(ii)} Oxx with $\y$ as input selects atoms in $\Qc$ during the first $g$ iterations and an atom $\a_i$, $i\notin \Qcs\backslash\Qc$ at the $(g+1)$th iteration.
  \end{lemma}

This result is of in interest in the analysis of Oxx at intermediate iterations since it shows that if $\mu<\frac{1}{2k-g-1}$ is not satisfied, it exists scenarios where Oxx selects good atoms during the first $g$ iterations and then fails at the subsequent step. 
%Finally, we emphasize that the RIC-based condition \eqref{eq:RICCOMPQ} derived in \cite{Karahanoglu2012Theoretical} also enjoys a type of worst-case necessity. In particular, the following result shows that \eqref{eq:RICCOMPQ} is almost tight for the success of OMP$_\Qc$ in the following sense:
% \begin{lemma}\label{lem:quasi-tighnessRIC}
%There exists a dictionary $\A$, a $k$-term representation $\y$ and  a set $\Qc$ with $\vert\Qc\backslash\Qcs\vert=b$ and $\vert\Qc\cap\Qcs\vert=g$, 
%such  that: \emph{(i)} $\delta_{k+b+1}=\frac{1}{\sqrt{k-g}+1}$;
%  \emph{(ii)} OMP$_\Qc$ with $\y$ as input selects a bad atom at the first iteration.
%  \end{lemma}
%The proof of this lemma is reported to ....

\subsection{Results and state-of-the-art conditions for $(P_p)$ and $(P_{p,\Qc})$}
\label{ssec:ContribPpQ}
The performance associated to the resolution  of  $(P_p)$ has been widely studied during the last decade. 
%The contributions of the literature includes uniform conditions guaranteeing the success of $(P_p)$ for any $k$-sparse vector, see \eg \cite{Fuchs2004Sparse,Gribonval2003Sparse},  and probabilistic analyses ensuring the success with high probability for some families of random measurements, see \eg \cite{Donoho2005Neighborliness,Wainwright2009Sharp}. We focus hereafter on the former group of contributions.
 Among the noticeable works dealing with uniform and (worst-case) necessary conditions, one can first mention the seminal paper by Fuchs \cite{Fuchs2004Sparse} in which the author showed that the success of $(P_1)$ only depends on the sign of the nonzero components in $\xs$. More recently, Wang \etal provided in \cite{Wang2011Performance} sufficient and worst-case necessary conditions for the success of $(P_p)$, with $p\in(0,1)$, depending on the sign-pattern of $\xs$. On the other hand, Gribonval\&Nielsen derived in \cite{Gribonval2003Sparse} the ``Null-Space Property", a tight conditions for the recovery of any $k$-sparse vector via $(P_p)$.
%  is due to Gribonval\&Nielsen in \cite{Gribonval2003Sparse}: 
%\begin{theorem}[Null-space property {\cite[Lemma 1]{Gribonval2003Sparse}}]\label{th:NSP}
%Assume $\spark(\A)>k$ and let
%\begin{align}
%\theta_p(k)\triangleq \max_{\vert \Qcs \vert = k} \max_{\v\in\ker_0(\A)} \Bigl\{ \frac{\| \v_\Qcs \|_p^p}{\| \v_{{\bar{\Qc}}^\star}\|_p^p} \Bigr\}. \label{eq:deftheta}
%\end{align}
%For any $p\in [0,1]$, if 
%\begin{align}
%\theta_p(k)\leq1 \label{eq:NSP}
%\end{align}
%then any $k$-sparse vector $\xs$ is a minimizer of \eqref{eq:SRproblem_rel} with $\y=\A\xs$ as input. Moreover, if the inequality in \eqref{eq:NSP} holds strictly, $\xs$ is the unique minimizer of \eqref{eq:SRproblem_rel}. 
%
%Conversely, if \eqref{eq:NSP} is not satisfied, there exists a $k$-sparse vector $\xs$ such that $\xs$ is not a minimizer of \eqref{eq:SRproblem_rel} with $\y=\A\xs$ as input.
%\end{theorem}
%%
%%Condition \eqref{eq:NSP} is usually referred to as ``Null-Space Property" (NSP) and has been rediscovered many times in the literature, see \eg \cite{Cohen2009Compressed, Hassibi}. 
%
%Interestingly, let us note that a condition similar to \eqref{eq:NSP} hold for the recovery of a particular support $\Qcs$ if we remove the left-most maximization in \eqref{eq:deftheta}, see \cite{Gribonval2003Sparse}.

%Unfortunately, although being tight, the Null-Space Property (NSP) turns out to be impractical in many situations,  the evaluation of $\theta_p(k)$ being indeed of combinatorial complexity. As a consequence, several conditions, easier to evaluate, based on RIC and mutual coherence have been derived in the literature.
Other conditions, easier to evaluate, have also been proposed in terms of RIC and mutual coherence. On the one hand, the use of RIC-based conditions was ignited by Candes, Romberg and Tao in their seminal work  \cite{Candes2005Decoding}. Candes refined this result in \cite{Candes2008Restricted} and some improvements were proposed by others authors in \cite{Foucart2009Sparsest, Cai2010Shifting}.

%On the one hand, the use of RIC-based conditions was ignited by Candes, Romberg and Tao in their seminal work  \cite{Candes2005Decoding}. In this contribution, the authors provided a sufficient condition of success for $(P_1)$ in terms of $(a+1)k$-order RIC, with $a>1$. 
%Candes refined this result in \cite{Candes2008Restricted} and showed that
%\begin{align}
%\delta_{2k}<\left(1+\sqrt{2}\right)^{-1} \label{eq:RICP1}
%\end{align}
%is sufficient for the success of $(P_1)$. In \cite{Foucart2009Sparsest, Cai2010Shifting}, some improvements of the constant appearing in the right-hand side of \eqref{eq:RICP1} were  proposed.  Finally, worst-case necessary conditions in terms of RICs were recently discussed in \cite{Davies2009Restricted}. 

On the other hand, guarantees for $(P_0)$ and $(P_1)$ based on the mutual coherence were early proposed in \cite{Donoho2001Uncertainty} for the particular setup of sparse representations in a union of two orthogonal bases. Several authors later proved independently that condition \eqref{eq:CBcondition} ensures the success of $(P_0)$ and $(P_1)$ for any $k$-sparse vector in arbitrary redundant dictionaries, see \eg \cite{Fuchs2004Sparse, Gribonval2003Sparse}. This condition was then shown to be valid for the success of $(P_p)$ with $p\in[0,1]$ in \cite{Gribonval2007Highly}. Finally, Cai\&Wang emphasized in \cite[Th. 
3.1]{Cai2010Stable}   that \eqref{eq:CBcondition} is also worst-case necessary (in some sense) for the success of $(P_p)$. 

 Recently, several authors took a look at conditions ensuring the success of \eqref{eq:SRproblemp_Q} when some partial information $\Qc$ is available about the support $\Qcs$. %All the contributions that we have identified in this vein focus on the case $p=1$. First, 
First, a ``truncated" NSP  generalizing the standard NSP has been derived in \cite[Th. 2.1]{Bandeira2011Partially}, \cite[Th. 3.1]{Wang2010Sparse}  and  \cite[Th. 3.1]{Chen2010Convergence}:
 %has been derived  in two independent contributions, see \cite[Theorem 2.1]{Bandeira2011Partially} and \cite[Theorem 3.1]{Wang2010Sparse}. This type of condition was also shown to hold for \eqref{eq:SRproblemp_Q}, $p\in[0,1]$, in \cite{Chen2010Convergence}. We summarize these results in the following theorem:
 \begin{theorem}[Truncated NSP]\label{th:nestingpropertypartial}
Assume that $\spark(\A)>k+b$ and let
\begin{align}
\theta_p(k,g,b)\triangleq \max_{\vert \Qcs \vert = k} \max_{\substack{\vert \bar{\Qc}^\star\cap\Qc \vert = b\\ \vert \Qcs\cap\Qc \vert = g}} \max_{\v\in\ker_0(\A)} \biggl\{ \frac{\| \v_{\Qcs\backslash\Qc} \|_p^p}{\| \v_{\overline{\Qc\cup\Qcs}}\|_p^p} \biggr\}. \label{eq:defthetapartial}
\end{align}
For any $p\in [0,1]$, if 
\begin{align}
\theta_p(k,g,b)\leq1, \label{eq:NSPpartial}
\end{align}
then any $k$-sparse vector $\xs$ is a minimizer of \eqref{eq:SRproblemp_Q} for any partial support estimate $\Qc$ such that  $\vert\Qcs\cap\Qc\vert=g$ and $\vert\bar{\Qc}^\star\cap\Qc\vert=b$. Moreover, if the inequality in \eqref{eq:NSPpartial} holds strictly, $\xs$ is the unique solution of \eqref{eq:SRproblemp_Q}. Conversely, if \eqref{eq:NSPpartial} is not satisfied, there exists a $k$-sparse vector $\xs$ and a support estimate $\Qc$ satisfying  $\vert\Qcs\cap\Qc\vert=g$ and $\vert\bar{\Qc}^\star\cap\Qc\vert=b$, such that $\xs$ is not a  minimizer of \eqref{eq:SRproblemp_Q} with $\y=\A\xs$ as input.
\end{theorem}
We note that the denominator in the right-hand side of \eqref{eq:defthetapartial} is always non-zero because of the hypothesis $\spark(\A)>k+b=\vert\Qcs\cup\Qc\vert$ (see also Appendix \ref{app:ProofResultsPp}). 
The direct part of Theorem \ref{th:nestingpropertypartial} is proved in \cite{Wang2010Sparse} and \cite{Chen2010Convergence} for $(P_{1,\Qc})$ and \eqref{eq:SRproblemp_Q}, respectively. In \cite{Bandeira2011Partially}, the authors demonstrated both the direct and  converse parts of Theorem \ref{th:nestingpropertypartial}  for $(P_{1,\Qc})$. We verified that the converse part  of Theorem \ref{th:nestingpropertypartial} also holds for \eqref{eq:SRproblemp_Q}, $p\in[0,1]$. The proof is  very similar to the exposition in \cite{Bandeira2011Partially} and \cite{Gribonval2003Sparse},  and is therefore not reported here. We note that Theorem \ref{th:nestingpropertypartial} reduces to the standard NSP  as soon as $g=b=0$. 
 
Several authors also proposed recovery guarantees in terms of RICs, see \cite{Vaswani2010ModifiedCS,Jacques2010Short,Friedlander2012Recovering}. In \cite{Vaswani2010ModifiedCS}, the authors identified a sufficient condition for the success of $(P_{1,\Qc})$ and show that the latter condition is weaker than a condition derived in \cite{Candes2005Decoding} for the non-informed setting as long as $\Qc$ contains a ``sufficiently" large number of good atoms. This result was later extended by Jacques \cite{Jacques2010Short} to the cases of compressible signals and noisy observations. Finally, in \cite{Friedlander2012Recovering}, Friedlander \etal generalized the RIC condition derived in \cite{Candes2005Decoding,Candes2008Restricted} to the partially-informed paradigm considered in this paper. In particular, the authors showed that the following condition\footnote{We have adapted the formulation of the condition derived in \cite{Friedlander2012Recovering} to the particular setup and notations considered in this paper.}  
 \begin{align}
\delta_{2k}<\Biggl(1+\sqrt{2 \left(1+\frac{b-g}{k}\right)}\Biggr)^{-1}, \label{eq:RICP2}
\end{align}
is sufficient for the success of \eqref{eq:SRproblemp_Q}. Interestingly, if $g=b=0$, one recovers the standard condition $\delta_{2k}<(1+\sqrt{2})^{-1}$ by Candes for the success of $(P_1)$, \cite{Candes2008Restricted}. Finally, we also mention the work by Khajehnejad \etal  \cite{AminKhajehnejad2009Weighted} where a Grassman angle approach was used to characterize a class of signal which can be recovered by (a variant of) \eqref{eq:SRproblem_Q}.
%Clearly, \eqref{eq:RICP2} is a weaker condition than \eqref{eq:RICP1} as soon as $g>b$. Moreover, one recovers Candes condition as soon as $b=g$. 

% In this paper, we emphasize that tight conditions of success for $(P_{p,\Qc})$ can be derived in terms of  ``partial" NSP for \emph{any} $p\in[0,1]$. More specifically, the following result holds:

%Let us notice that the denominator in \eqref{th:nestingpropertypartial} is well-defined for any $\v\in\ker_0(\A)$ since we assumed by hypothesis that $\spark(\A)>k+b$. 
%The case $p=1$ of this theorem corresponds to the results stated in \cite{Bandeira2011Partially,Wang2010Sparse}. Interestingly, one recovers the standard NSP of Theorem \ref{ssec:ContribPpQ} as soon as $g=0$ and $b=0$.

In this paper, we will show that the quantities $\theta_p(k,g,b)$ involved in the truncated NSP obey an ordering property and can be related to the partial ERC stated in Theorem \ref{th:SoussenERC} (see Theorems \ref{th:nestingproperty} and \ref{th:nesting_ERCNSP} below). As a consequence of these results, together with Theorem \ref{th:mainth}, we obtain that a coherence-based condition, similar to the one obtained for Oxx$_\Qc$, holds for the success of \eqref{eq:SRproblemp_Q}:
 
%We show moreover, that these generalized NSP conditions enjoy a nesting property similar to the one stated in Theorem \ref{th:nestingproperty}.Moreover, the following nesting property holds:
%\begin{align}
%%\theta_0(k,g,b)\leq \theta_q(k,g,b)\leq \theta_p(k,g,b) \leq \theta_1(k,g,b)\leq \theta_{OMP}(k,g,b). \label{eq:nestingpartialprop}
%\theta_0(k,g,b)\leq \theta_q(k,g,b)\leq \theta_p(k,g,b) \leq \theta_1(k,g,b). \label{eq:nestingpartialprop}
%\end{align}
%for any $p,q$  such that $0\leq q \leq p\leq 1$. 

%It is interesting to note from \eqref{eq:nestingpartialprop} that the  quantities involved in the extended NSP, that is $\theta_p(k,g,b)$, is related to the partial ERC for OMP derived in \cite{}. 

%The NSP condition suffers from the same computational burden as the ERC. Hence, conditions based on the 

\begin{theorem}[$\mu$-based uniform condition for \eqref{eq:SRproblemp_Q}]\label{th:mainthPp}
  Consider a $k$-term representation $\y
    =\A\xs$ and  a support $\Qc$ such that $\vert\Qcs\cap\Qc\vert=g$ and $\vert\bar{\Qc}^\star\cap\Qc\vert=b$. If $\mu<\frac{1}{2k-g+b-1}$ holds, 
%\begin{align}
%\mu < \frac{1}{2k-g+b-1}, \label{eq:mainBoundPpQ}
%\end{align}
then $\xs$ is the unique minimizer of \eqref{eq:SRproblemp_Q}. 
Conversely, there exists a dictionary $\A$ and a $k$-term
  representation $\y=\A\xs$ such that: \emph{(i)} $\mu=\frac{1}{2k-g+b-1}$;
  \emph{(ii)} $\xs$ is not the unique minimizer of  \eqref{eq:SRproblemp_Q}.
\end{theorem}

The direct part of Theorem \ref{th:mainthPp} is proved in section \ref{sec:proofsPp}. The converse part will be shown in section \ref{sec:NC}. % together Theorem \ref{} will be proved in ....

 Interestingly, similar to the result by Friedlander \etal in \eqref{eq:RICP2}, we can notice that our coherence-based condition becomes weaker than its standard counterpart \eqref{eq:CBcondition} as soon as $g>b$, that is, when at least $50\%$ of the atoms of $\Qc$ belongs to $\Qcs$. In other words, the success of \eqref{eq:SRproblemp_Q} is ensured under conditions less restrictive than for $(P_p)$ as soon as  $\Qc$ provides a ``sufficiently reliable" information about $\Qcs$. %We will however emphasize that our coherence-based condition and \eqref{eq:RICP2} enjoys some form of complementarity in the sense where none of the condition implies the other

\subsection{Relationships between conditions for $(P_{p,\Qc})$ and Oxx$_{\Qc}$}
\label{ssec:Relationship}
In this section, we discuss the implications (or non-implications) existing between some of the conditions mentioned above. 
 First, we emphasize that an ordering property, similar to the one derived in Gribonval\&Nielsen in  \cite[Lemma 7]{Gribonval2007Highly} for $(P_p)$, still holds for the truncated NSPs defined in Theorem \ref{th:nestingpropertypartial}:
\begin{theorem}[Ordering Property of Truncated  NSPs]\label{th:nestingproperty}
If $0\leq q \leq p\leq 1$ and $\spark(\A)>k+b$, the following nesting property holds:
%\theta_0(k)\leq \theta_q(k)\leq \theta_p(k) \leq \theta_1(k)\leq \theta_{Oxx}(k,0,0).
%\theta_0(k)\leq \theta_q(k)\leq \theta_p(k) \leq \theta_1(k).
\begin{align}
%\theta_0(k,g,b)\leq \theta_q(k,g,b)\leq \theta_p(k,g,b) \leq \theta_1(k,g,b)\leq \theta_{OMP}(k,g,b). \label{eq:nestingpartialprop}
\theta_0(k,g,b)\leq \theta_q(k,g,b)\leq \theta_p(k,g,b) \leq \theta_1(k,g,b). \label{eq:nestingpartialprop}
\end{align}
\end{theorem}
%The last inequality is from Tropp in \cite[]{Tropp2004Greed}. 
The proof of this result is reported to section~\ref{sec:proofsPp}. Clearly, one recovers Gribonval\&Nielsen's ordering property as a particular case of \eqref{eq:nestingpartialprop} as soon as $g=b=0$. This ordering property implies that any uniform condition for $(P_{p,\Qc})$ is also a sufficient condition of success for $(P_{q,\Qc})$ with $q\in[0,p]$. In particular, the guarantees derived in \cite{Vaswani2010ModifiedCS,Jacques2010Short,Friedlander2012Recovering} for $(P_{1,\Qc})$ also ensure the success of $(P_{p,\Qc})$ for 
$p\in[0,1]$.

Secondly, we show that the truncated NSPs share some connections with the partial ERC for OMP defined in \eqref{eq:ERCunif}. Specifically, we have
\begin{theorem} \label{th:nesting_ERCNSP} If $\spark(\A)>k+b$, then
%If $k, g, b$ are such that $\A_{\Qcs\cup\Qc}$ is full-rank $\forall\Qcs,\Qc$ with $\vert\Qc\cap\Qcs\vert=g$ and $\vert\Qc\backslash\Qcs\vert=b, 
\begin{align}
\theta_1(k,g,b)\leq \theta_{\mathrm{OMP}}(k,g,b).\label{eq:orderingL1OMP}
\end{align}
\end{theorem}
The proof of this result is reported to section~\ref{sec:proofsPp}.  
 This inclusion generalizes Tropp's result \cite[Th. 3.3]{Tropp2004Greed} to the paradigm of sparse representation with partial support information, namely ERC-OMP is a sufficient condition of success for $(P_{1,\Qc})$ (and thus for any $(P_{p,\Qc})$ with $p\in[0,1]$ by virtue of Theorems \ref{th:nestingpropertypartial} and \ref{th:nestingproperty}). As an important by-product of this observation, it turns out that any uniform guarantee of success for OMP$_\Qc$ is also a sufficient condition of success for \eqref{eq:SRproblemp_Q}.

It is noticeable that an  ordering similar to \eqref{eq:orderingL1OMP} does not generally hold between $\theta_1(k,g,b)$ and $\theta_{\mathrm{OLS}}(k,g,b)$ for all $k, g, b$. Indeed, on the one hand, $\theta_{\mathrm{OLS}}(k,0,0)\geq\theta_1(k,0,0)$ since $\theta_{\mathrm{OLS}}(k,0,0)=\theta_{\mathrm{OMP}}(k,0,0)$.
On the other hand, we exhibit a case for which $\theta_{\mathrm{OLS}}(k,g,b)<1<\theta_{1}(k,g,b)$:

% This is emphasized in the following example:

\begin{example}
In this example, we construct a dictionary such that 
\begin{align}
\theta_1(k,g,b) >1,\label{eq:ex1_noNSP}\\
\theta_{\mathrm{OLS}}(k,g,b)<1, \label{eq:ex1_ERCOLS}
\end{align}
%$\theta_1(k,g,b) >1$ and $\theta_{OLS}(k,g,b)<1$ 
for some $k, g, b$. %We consider the case where $k=2$, $g=1$, $b=0$. 
Let $n\geq 3$ and define the matrix
%\begin{align}
%\G &=
%\left(
%\begin{array}{ccc}
%1 & \alpha & \beta \oneb_{n-2}^T\\
%\alpha & 1& \beta \oneb_{n-2}^T\\
%\beta \oneb_{n-2} & \beta \oneb_{n-2} & \I_{n-2}
%\end{array}
%\right), \qquad \mbox{for $\alpha$, $\gamma\in\R$,}
%\end{align}
\begin{align}
\G &=
\left(
\begin{array}{ccc}
\I_{n-2} & \beta \oneb_{n-2} & \beta \oneb_{n-2}\\
\beta \oneb_{n-2}^T & 1& \alpha\\
\beta \oneb_{n-2}^T& \alpha & 1
\end{array}
\right), \qquad \mbox{for $\alpha$, $\beta\in\R$,}
\end{align}
which will play the role of the Gram matrix of the dictionary, that is $\G=\A^T\A$.  Since $\G$ is symmetric it allows for the following eigenvalue decomposition:
\begin{align}
\G = \U \Diag \U^T,
\end{align}
where $\U$ (resp. $\Diag$) is the unitary matrix whose columns are the eigenvectors (resp. the diagonal matrix of
eigenvalues) of $\G$. Letting
\begin{align}
\alpha &= \frac{1}{2}\gamma^2(n-2)-1,\\
\beta   &= -\frac{\gamma}{2},
\end{align}
with 
\begin{align}
\vert \gamma\vert<(n-2)^{-1}, \label{eq:defgamma}
\end{align}
it is easy to see that $\G$ is a semi-definite positive matrix with one single zero eigenvalue.
%\begin{align}
%\vert \gamma\vert<\frac{2}{\sqrt{2(n-2)-1}}.
%\end{align}
%\begin{align}
%\vert \gamma\vert<\frac{1}{n-2}.
%\end{align}
The zero eigenvalue is located in the lower-right corner of $\Lambda$ and the corresponding eigenvector writes, up to a normalization factor, as
\begin{align}
\begin{array}{ccc}
\v = [\gamma\oneb_{n-2}^T& 1& 1]^T. \label{eq:eigenvectorA}
\end{array}
\end{align}
We define $\A\in\R^{n-1\times n}$ as 
\begin{align}
\A &= \Upsilon\U^{T},
\end{align}
where $\Upsilon \in \mathbb{R}^{n-1\times n}$ is such that
\begin{align}
\Upsilon(i,j) = \left\{
\begin{array}{cl}
\sqrt{\Lambda(i,i)} & \mbox{if $i=j$,}\\
0 & \mbox{otherwise.}
\end{array}
\right.
\end{align}
Hence, $\Upsilon^T  \Upsilon=\Lambda$ and $\A^T \A = \U \Upsilon^T  \Upsilon \U^T = \U \Diag \U^T = \G$.

Now, $\A$ is such that \eqref{eq:ex1_noNSP} and \eqref{eq:ex1_ERCOLS} hold for $k=2$, $g=1$, $b=0$. %$\theta_1(2,1,0) >1$ and $\theta_{OLS}(2,1,0)<1$.
Indeed, on the one hand it can easily be seen that $\ker(\A)=\ker(\G)$
and $\ker(\A)$ corresponds therefore to the one-dimensional subspace
defined by $\v$ in \eqref{eq:eigenvectorA}. This implies that
$\spark(\A)=n$. Moreover, considering $\Qcs=\{n-1,n\}$ and
$\Qc=\{n-1\}$, we have
\begin{align}
\theta_1(2,1,0)\geq \frac{\|\v_{\Qcs\backslash\Qc}\|_1}{\|\v_{\overline{\Qcs\cup\Qc}}\|_1} =\frac{1}{(n-2)\gamma}>1
\end{align}
where the first inequality follows from the definition of $\theta_1(2,1,0)$ and the last one from the fact that $\gamma<(n-2)^{-1}$. 

On the other hand, since $\spark(\A)=n\geq k+1$ and there is
  only one atom in $\Qcs\backslash\Qc$, we have from \cite[Th.
6]{Soussen2013Joint} that necessarily
$\theta_{\mathrm{OLS}}(2,1,0)<1$.
\end{example}

In the previous example, we provided a simple scenario where OLS$_\Qc$ succeeds in recovering $\x^\star$ when $g=k-1$. It can also be observed that $(P_{\Qc,0})$ also succeeds in this particular example. Indeed, using the definition of $\v$ in \eqref{eq:eigenvectorA}, we have
\begin{align}
\theta_0(2,1,0) =\frac{1}{(n-2)}<1.
\end{align}
More generally, it can be shown that there exists an equivalence between the success of $(P_{\Qc,0})$  and OLS$_\Qc$ when $g=k-1$. This follows from the fact that the problem resolved by OLS$_\Qc$ when $g=k-1$, that is \eqref{eq:atomselection1}, is exactly equivalent to $(P_{\Qc,0})$. From a more technical point of view, it can easily be seen that condition $\theta_0(k,k-1,b)<1$ can be rephrased as
\begin{align}
k+b+1<\spark (\A).
\end{align}
Now, by slightly extending the arguments developed in \cite[Th. 6]{Soussen2013Joint}, the latter condition is also sufficient and worst-case necessary for the success of  OLS$_\Qc$ when $g=k-1$. This observation thus demonstrates the optimality of OLS$_\Qc$ when the informed support contains all the correct atoms but one.

%From Theorems \ref{th:nestingproperty} and \ref{th:nesting_ERCNSP}, the inclusion of the condition ruling the worst-case performance of $(P_p)$ and OMP$_{\Qc}$. 

%In Theorems \ref{th:mainth} and \ref{th:mainthPp}, we derived a novel guarantee of success for $(P_{p,\Qc})$ and Oxx$_{\Qc}$ in terms of mutual coherence of the dictionary. On the other hand, other conditions were previously proposed in terms of RICs, see \eqref{eq:RICCOMPQ} for OMP$_{\Qc}$ and \eqref{eq:RICP2} for $(P_{1,\Qc})$. Hence, one legitimate question arises: is there any implication from \eqref{eq:CBcondition_partial} to \eqref{eq:RICCOMPQ} or vice-versa? 
%%The answer to this question is negative as shown herefetin the following result:
%%The next potentially interesting question is: ``Is there any implication between \eqref{eq:mainBound} and \eqref{eq:RICCOMPQ}?".  
% We show hereafter that the answer to this question is negative. 
%In particular, we emphasize two particular instances of dictionary such that \eqref{eq:CBcondition_partial} is satisfied but \eqref{eq:RICCOMPQ} is not, and vice versa. We construct our examples in the case where $g=0$ for conciseness. Similar construction can however be applied to derive examples in the general case.
%
%QUESTION: SAME TYPE OF RESULT FOR \eqref{eq:RICP2}?
\subsection{Non-implication between the mutual and RIP conditions for Oxx$_\Qc$}\label{ssec:complRICmu}

In Theorem \ref{th:mainth}, we derived a novel guarantee of success for Oxx$_{\Qc}$ in terms of mutual coherence of the dictionary. On the other hand, other conditions were previously proposed in terms of RICs, see \eqref{eq:RICCOMPQ} for OMP$_{\Qc}$. Hence, one legitimate question arises: is there any implication from \eqref{eq:CBcondition_partial} to \eqref{eq:RICCOMPQ} or vice-versa? 
%%The answer to this question is negative as shown herefetin the following result:
%%The next potentially interesting question is: ``Is there any implication between \eqref{eq:mainBound} and \eqref{eq:RICCOMPQ}?".  
 We show hereafter that the answer to this question is negative. 
In particular, we exhibit two particular instances of dictionary such that \eqref{eq:CBcondition_partial} is satisfied but \eqref{eq:RICCOMPQ} is not, and vice versa. We construct our examples in the case where $b=0$ for the sake of conciseness. Similar constructions can however be applied to derive examples in the general case.
%
%QUESTION: SAME TYPE OF RESULT FOR \eqref{eq:RICP2}?

%A similar construction can be carried out to show the non-implication of \eqref{eq:mainBound} and \eqref{eq:RICP2}. The details are however omitted for the sake of conciseness. Moreover, our examples are constructed in the particular case $b=g=0$. However, the same type of result can straightforwardly be applied to the more general case (TO CHECK). 

\begin{example}[$\A$ satisfies \eqref{eq:CBcondition_partial} but not \eqref{eq:RICCOMPQ}] \label{exple:RICmu1}
%Let us first focus on the case $b=0$.
 Let us consider $\A\in\R^{(k+1)\times (k+1)}$ such that
\begin{align}
\G \triangleq \A^T \A
% = \left(
%\begin{array}{ccccc}
%1 & -\mu & \ldots & \ldots & -\mu\\
%-\mu & 1 & -\mu & &
%\end{array}
%\right)
\end{align}
and
\begin{align}
G_{i,j}&=\left\lbrace
\begin{array}{cl}
-\mu & i\neq j\\
1 & i=j
\end{array}
\right.
\end{align}
with $\mu\leq1/k$. 
We have therefore
\begin{align}
\lambda_{max}(\G)&=1+\mu \qquad\mbox{(with multiplicity $k$)},\\
\lambda_{min}(\G)&=1- k\mu,
\end{align}
and 
\begin{align}
\delta_{k+1}
&=\max \{1-\lambda_{min}(\G),\lambda_{max}(\G)-1\}\nonumber\\
&=k\mu.
\end{align}
We can freely set $\mu=\alpha/(2k-g-1)$ with $0\leq g<k$ and $\alpha\in(0,1)$
since this yields $\mu<1/k$. Then, $\mu$ trivially satisfies
\eqref{eq:CBcondition_partial}. On the other hand, $\delta_{k+1}$ can
be written as
%\remCS{ajout le dernier $\geq$}
\begin{align}
\delta_{k+1} = \frac{\alpha k}{2k-g-1}\geq \alpha/2.
\label{eq:delta_k+1}
\end{align}
For any  $g<k-1$, there exist $\alpha\in(0,1)$ and $k$ such that \eqref{eq:RICCOMPQ} is not verified.
For example, for $k$ sufficiently large and fixed $g<k-1$,
$\delta_{k+1}$ in \eqref{eq:delta_k+1} does not satisfy \eqref{eq:RICCOMPQ} since 
the right-hand side of~\eqref{eq:RICCOMPQ} tends towards
0 when $k$ tends to infinity. 
%The same type of result can be obtained for $b\neq 0$ by considering the dictionary $\check{\A}\in \R^{k+b+1\times k+b+1}$ defined as
%\begin{align}
%\check{\A} &=
%\left[
%\begin{array}{cc}
%\A & \mathbf{0}_{k+1\times b}\\
%\mathbf{0}_{k+1\times b}^T & \I_{b\times b}.
%\end{array}
%\right]
%\end{align}
%where $\A$ is such that \eqref{} is satisfied. 
\end{example}

\begin{example}[$\A$ satisfies \eqref{eq:RICCOMPQ} but not \eqref{eq:CBcondition_partial}]
Let 
\begin{align}
\A &\triangleq\left(
\begin{array}{ccc}
\a_1 & \a_2 & \H
\end{array}
\right)\in \R^{(k+1)\times (k+1)}
\end{align}
be such that
\begin{align}
 \a_1^T \a_2 &=\mu,\\
\H^T \a_1 &= \H^T \a_2 = \mathbf{0},\\
\H^T\H&=\I_{k-1}.
\end{align}
Then, we easily have
\begin{align}
\lambda_{max}(\G)&=1+\mu,\\
\lambda_{min}(\G)&=1- \mu,
\end{align}
and 
\begin{align}
\delta_{k+1}
&=\max \{1-\lambda_{min}(\G),\lambda_{max}(\G)-1\}\nonumber\\
&=\mu.
\end{align}
Let us set
$\delta_{k+1}=\mu=\alpha/(\sqrt{k-g}+1)$ with $\alpha\in(0,1)$. Then,
$\delta_{k+1}$ trivially satisfies \eqref{eq:RICCOMPQ}. On the other
hand, $\mu>1/(2k-g-1)$ holds for sufficiently large $k$ 
and a fixed value of $g<k$.
\end{example}

Finally, we mention that, following the same procedures as above, one can derive examples for which \eqref{eq:RICP2} is satisfied but \eqref{eq:CBcondition_partial} is not for some value of $k,g,b$, and vice-versa. The details are however not reported here for the sake of conciseness.  
%Finally, in order to complete our picture of the relationships existing between the conditions ensuring the success of sparse representation algorithms with partial support information, we emphasize that the RIC-based sufficient condition \eqref{eq:RICCOMPQ} for $OMP_{\Qc}$ is almost tight in the following sense:
%\begin{theorem} \label{th:RICOMPQtight} There exists a dictionary $\A$, a $k$-term representation $\y=\A\xs$ and a support estimate $\Qc$ with $\vert\Qc\backslash\Qcs\vert=b$ and $\vert\Qc\cap\Qcs\vert=g$, such that: (i) $\delta_{k+b+1}=\frac{1}{\sqrt{k-g}+1} $; (ii) OMP$_\Qc$ selects a bad atom at the first iteration.
%\end{theorem}

\section{Sufficiency of \eqref{eq:CBcondition_partial} for OMP$_\Qc$}\label{sec:SCOMP}

In this section, we prove the direct part
of Theorem \ref{th:mainth} for OMP$_\Qc$. The result is a direct
consequence of Proposition \ref{th:ubERC} stated below, which provides
an upper bound on the left-hand side of \eqref{eq:SoussenERC} only
depending on the coherence of the dictionary $\A$:
\begin{proposition}\label{th:ubERC}
Let $\Qcs$ and $\Qc$ be such that $\vert \Qcs\vert=k$, $\vert \Qcs\cap\Qc\vert=g$ and $\vert \bar{\Qc}^\star\cap\Qc\vert=b$. If 
\begin{align}
\mu< \frac{1}{k+b-1},
\end{align}
then
\begin{align}
\max_{i\notin \Qcs} \| \tA_{\Qc^\star \backslash \Qc}^\dag \tilde{\a}_{i} \|_1\leq \frac{(k-g)\mu}{1-(k+b-1)\mu}. \label{eq:upperboundERC}
\end{align}
\end{proposition}
The sufficient condition for OMP$_\Qc$ stated in Theorem \ref{th:mainth} then derives from Proposition \ref{th:ubERC} and Theorem \ref{th:SoussenERC}. Indeed, we see from Proposition \ref{th:ubERC} that
\begin{align}\label{eq:suffcondmu}
\frac{(k-g)\mu}{1-(k+b-1)\mu}<1
\end{align}
implies \eqref{eq:SoussenERC}. Moreover, by reorganizing the latter expression, it is easy to see that \eqref{eq:suffcondmu} is equivalent to  \eqref{eq:CBcondition_partial}. To prove Theorem \ref{th:mainth} it thus remains to apply Theorem~\ref{th:SoussenERC}. 
Now, the  full-rankness of $\A_{\Qcs\cup\Qc}$ in the hypotheses  of 
Theorem~\ref{th:SoussenERC} is implicitly enforced by
\eqref{eq:CBcondition_partial}. Indeed, as shown in \cite[Lemma~2.3]{Tropp2004Greed},
\begin{align}
\mu< \frac{1}{k+b-1}
\end{align}
implies that $\A_{\Qcs\cup\Qc}$ is full rank whenever $\vert
\Qcs\cup\Qc\vert=k+b$.  Hence, since $k+b-1<2k-g+b-1$,
\eqref{eq:CBcondition_partial} in turn implies that any
submatrix $\A_{\Qcs\cup\Qc}$ with $\vert \Qcs\cup\Qc\vert=k+b$ is full
rank.  Then, applying Theorem~\ref{th:SoussenERC},  we have that \eqref{eq:CBcondition_partial} is sufficient for the success of OMP$_\Qc$ in $k-g$ iterations.\\

Before proving Proposition \ref{th:ubERC}, we need to define some
quantities characterizing the \emph{projected} dictionary $\tA$
appearing in the implementation of OMP (see \eqref{eq:atomselection2})
and state some useful propositions.
In the following definition, we generalize the concept of restricted
isometry property (RIP) \cite{Candes2005Decoding} to projected
dictionaries, under the name projected RIP (P-RIP):
\begin{defi}\label{def:GRIC}
 Dictionary $\A$ satisfies the P-RIP($\ld_{q,l}$,$\ud_{q,l}$) if and only if $\forall \Qc',
  \Qc$ with $\vert \Qc'\vert=q$, $\vert\Qc\vert=l$, $\Qc \cap
  \Qc'=\emptyset$, $\forall \x_{\Qc'}$ we have
\begin{align}
(1-\ld_{q,l}) \| \x_{\Qc'} \|^2 \leq  \| \tA_{\Qc'}^{\Qc} \xQp \|^2 \leq (1+\ud_{q,l}) \| \x_{\Qc'} \|^2.
\end{align}
 %  $\ub(q,l)$ is said to a be a $(q,l)-$upper bound if $\forall \Qc',
%  \Qc$ with $\vert \Qc'\vert=q$, $\vert\Qc\vert=l$, $\Qc \cap
%  \Qc'=\emptyset$, $\forall \x_{\Qc'}$ we have
%  \begin{align}
%    \ub(q,l) \| \x_{\Qc'} \|^2 \geq \| \tA_{\Qc'}^{\Qc} \xQp \|^2.
%  \end{align}
%  Similarly, $\lb(q,l)\geq 0$ is said to a be a $(q,l)-$lower bound
%  if $\forall \Qc', \Qc$ with $\vert \Qc'\vert=q$, $\vert\Qc\vert=l$,
%  $\Qc \cap \Qc'=\emptyset$, $\forall \x_{\Qc'}$, we have
%  \begin{align}
%\lb(q,l) \| \x_{\Qc'} \|^2 \leq \| \tA_{\Qc'}^{\Qc} \xQp \|^2.
%\end{align}
\end{defi}
The definition of the standard (asymmetric)
restricted isometry constants corresponds to the tightest possible
bounds when $l=0$ (see \eg
\cite{Foucart2009Sparsest,Davies2009Restricted}). For $l\geq
  1$, $\ld_{q,l}$ and $\ud_{q,l}$ can be seen as (asymmetric)
\emph{bounds} on the restricted isometry constants of \emph{projected}
dictionaries. Note that $\ud_{q,l}$ might be negative
since the columns of $\tA$ are not normalized
($\|\ta_i^{\Qc}\|\leq 1$). Note also that many well-known
properties of the standard restricted isometry constants (see
\cite[Proposition 3.1]{Needel_ACMels08} for example) remain valid for
$\ld_{q,l}$ and
$\ud_{q,l}$. %The result stated in Theorem \ref{th:mainth} is a
            %consequence of the two following propositions:

The next lemma provides an upper bound on the left-hand side of \eqref{eq:SoussenERC} only depending on the P-RIP
  constants: %$\ld_{q,l}$ and $\ud_{q,l}$:  
\begin{lemma}\label{UBERC} %If $\lb(k-l,l)>0$, then 
%Let $\Qc \subset \Qcs$, with $\vert \Qc\vert=l$, $\vert \Qcs\vert=k$.
Let $\Qcs$ and $\Qc$ be such that $\vert \Qcs\vert=k$, $\vert \Qcs\cap\Qc\vert=g$ and $\vert \bar{\Qc}^\star\cap\Qc\vert=b$.
 If $\ld_{k-g,g+b}<1$, then 
\begin{align}
%\max_{j\notin \Qcs} \| \tA_{\Qc^\star \backslash \Qc}^\dag \tilde{\a}_{j} \|_1 < (\leq ?) (k-l)\, \frac{\ub(2,l)-\lb(2,l)}{2\lb(k-l,l)}.\label{eq:UBERC}
\max_{i\notin \Qcs} \| \tA_{\Qc^\star \backslash \Qc}^\dag \tilde{\a}_{i} \|_1 \leq (k-g)\, \frac{\ud_{2,g+b}+\ld_{2,g+b}}{2(1-\ld_{k-g,g+b})}.\label{eq:UBERC}
\end{align}
\end{lemma}
%As a consequence, if the right-hand side of \eqref{eq:UBERC} is upper bounded by one, Soussen \etal's ERC is satisfied. 
The proof of Lemma \ref{UBERC} is reported to Appendix
\ref{sec:SCOMP}. The next lemma provides some possible values
for  $\ld_{q,l}$ and $\ud_{q,l}$ as a function of the coherence of the
dictionary $\A$:

\begin{lemma} \label{prop:linkRICmu} 
If $\mu<1/(l-1)$, then $\A$ satisfies the P-RIP($\ld_{q,l}$,$\ud_{q,l}$) 
for any $q\geq 0$
with 
\begin{align}
\ud_{q,l} &= (q-1)\mu \label{eq:relRICmu1},\\
\ld_{q,l}  &= (q-1)\mu +\frac{\mu^2 q l}{1-(l-1)\mu}.
\label{eq:relRICmu2}
\end{align}
\end{lemma}
The proof of this result is reported to Appendix \ref{sec:SCOMP}. We are now ready to prove Proposition \ref{th:ubERC}:\vspace{0.2cm}

\begin{IEEEproof}
\emph{(Proposition \ref{th:ubERC})} 
%The proof consists in
We rewrite the right-hand side of \eqref{eq:UBERC} as a function of
$\mu$.
% by using Proposition \ref{prop:linkRICmu}.  
From Lemma \ref{prop:linkRICmu}, we have that $\A$ satisfies the
P-RIP($\ld_{q,l}$,$\ud_{q,l}$) with constants defined in
\eqref{eq:relRICmu1}-\eqref{eq:relRICmu2} as long as
\begin{align}
\mu< \frac{1}{l-1}.\label{eq:subcondition_th1}
\end{align} 
Now, we have $\mu<1/(k+b-1)$ by hypothesis, which implies $\mu<1/(g+b-1)$.
Thus, Lemma \ref{prop:linkRICmu} can be applied with $l=g+b$. 
Using \eqref{eq:relRICmu1} and \eqref{eq:relRICmu2}, we
  calculate that:
\begin{align}
\frac{\ud_{2,g+b}+\ld_{2,g+b}}{2}&=\mu+\frac{\mu^2(g+b)}{1-(g+b-1)\mu}\\
&= \frac{\mu(\mu+1)}{1-(g+b-1)\mu},\\
1-\ld_{k-g,g+b}&=1-(k-g-1)\mu-\frac{\mu^2(k-g)(g+b)}{1-(g+b-1)\mu}\\
&= \frac{1-(k+b-2)\mu-(k+b-1)\mu^2}{1-(g+b-1)\mu}\\
&=\frac{(\mu+1)(1-(k+b-1)\mu)}{1-(g+b-1)\mu}.\label{eq:1moinslambda}
\end{align}
Therefore, the ratio in the right-hand side of \eqref{eq:UBERC}
can be rewritten as
\begin{align}
\frac{\ud_{2,g+b}+\ld_{2,g+b}}{2(1-\ld_{k-g,g+b})}
&= \frac{\mu}{1-(k+b-1)\mu}.\label{eq:ratio_lambda}
\end{align}
According to~\eqref{eq:1moinslambda},
$\mu<1/(k+b-1)\leq 1/(g+b-1)$ implies that $1-\ld_{k-g,g+b}>0$. 
Lemma~\ref{UBERC} combined with~\eqref{eq:ratio_lambda} implies that~\eqref{eq:upperboundERC} 
is met.
% holds as soon as $1-\ld_{k-l,l}>0$. 
% Using \eqref{eq:relRICmu1} and \eqref{eq:relRICmu2}, the latter condition rewrites
% %\begin{align}
% %\frac{1-(k-2)\mu-(k-1)\mu^2}{1-(l-1)\mu}>0.\label{eq:subcondth1_2}
% %\end{align}
% %Since \eqref{eq:subcondition_th1} is assumed to hold, the denominator is positive and \eqref{eq:subcondth1_2} is equivalent to
% \begin{align}
% 1-(k-2)\mu-(k-1)\mu^2=
% (\mu+1)(1-\mu(k-1))>0. \label{eq:subcondth1_3b}
% \end{align}
% This inequality is satisfied since $\mu<1/(k-1)$ by hypothesis.
\vspace{0.3cm}
\end{IEEEproof}

%Before concluding this section, let us remark that unlike
%Theorem~\ref{th:SoussenERC}, Theorem \ref{th:mainth} does not
%(explicitly) require all $m\times (k+b)$-submatrices $\A_{\Qcs\cup\Qc}$ to be
%full rank. However, this condition is implicitly enforced by
%\eqref{eq:CBcondition_partial}. Indeed, as shown in \cite[Lemma~2.3]{Tropp2004Greed},
%\begin{align}
%\mu< \frac{1}{k+b-1}
%\end{align}
%implies that $\A_{\Qcs\cup\Qc}$ is full rank when\addCS{ever} $\vert
%\Qcs\cup\Qc\vert=k+b$.  Hence, since $k+b-1<2k-g+b-1$,
%\eqref{eq:CBcondition_partial} \addCS{in turn} implies that any
%submatrix $\A_{\Qcs\cup\Qc}$ with $\vert \Qcs\cup\Qc\vert=k+b$ is full
%rank.  \suppCS{Finally, we remark that the full rankness of
%  $\A_{\Qcs\cup\Qc}$ implies that the projected submatrices
%  $\tA_{\Qc^\star \backslash \Qc}$ involved in Proposition~ xxx
%  ref{th:ubERC} are also full rank~ xxx
%  cite[Corollary~3]{Soussen2013Joint}. }

\section{Sufficiency of \eqref{eq:CBcondition_partial} for OLS$_\Qc$}\label{sec:SCOLS}

We now prove the sufficient condition for OLS$_\Qc$ stated in
Theorem \ref{th:mainth}. The result is a consequence of Proposition
\ref{lem:connect_OLS} and Lemma \ref{prop:bound_mutilde} stated below.
We first need to introduce the coherence of the \emph{normalized}
projected dictionary $\tilde{\B}$:
%(denoted by $\tilde{\B}$ with $\tilde{\b}_i=\tilde{\a}_i/\|\tilde{\a}_i\|$ if $i\notin\Qc$, and \zerob otherwise). 
%
%
%
\begin{defi}[Coherence of the normalized projected
  dictionary]\label{def:coherencePD2}
\begin{align}
\mu^{OLS}_l = \max_{\vert \Qc \vert=l}\max_{i \neq j}  \vert \langle 
\tilde{\b}_i^{\Qc}, \tilde{\b}_j^{\Qc} \rangle\vert. 
\end{align}
\end{defi}
%
%The following two lemmas aim to upper bound $\mu^{OLS}_l$ and the
%term $ \max_{i\notin \Qc^\star \backslash \Qc} \| \tilde{\B}_{\Qc^\star \backslash \Qc}^\dag \tilde{\b}_{i}
%\|_1$ involved in \eqref{eq:SoussenERC}.
%

The following proposition gives a sufficient condition on $\mu^{OLS}_{g+b}$ under which \eqref{eq:SoussenERC} is satisfied:
\begin{proposition} \label{lem:connect_OLS}
%
%Let $\Qc\subset \Qcs$ with $\vert \Qc \vert=l$, $\vert \Qcs \vert=k$. 
%  Let $\Qc \subset \Qcs$, with $\vert \Qc\vert=l$, $\vert\Qcs\vert=k$. 
Let $\Qcs$ and $\Qc$ be such that $\vert \Qcs\vert=k$, $\vert \Qcs\cap\Qc\vert=g$ and $\vert \bar{\Qc}^\star\cap\Qc\vert=b$.
  Assume that $\A_{\Qcs\cup\Qc}$ is full rank. If
  $\mu^{OLS}_{g+b}<1/(2k-2g-1)$,
%  If 
%\begin{align}
%\mu^{OLS}_l<1/(2k-2l-1), 
%\end{align}
 then 
 \begin{align}\label{eq:ERCOLS}
\max_{i\notin \Qc^\star} \| \tilde{\B}_{\Qc^\star
    \backslash \Qc}^\dag \tilde{\b}_{i} \|_1<1. 
\end{align}
\end{proposition}
\begin{IEEEproof}
When $\tilde{\b}_{i}=\mathbf{0}$, the result is obvious.
    When $\tilde{\b}_{i}\ne\mathbf{0}$, apply~\cite[Corollary
  3.6]{Tropp2004Greed} (that is: if $\A$ has normalized
  columns and $\mu<1/(2k-1)$ then Tropp's ERC is satisfied, \ie
  $\forall \Qc^\star$ such that $\vert \Qc^\star\vert=k$,
  $\max_{i\notin \Qcs} \|\A_{\Qc^\star}^\dag {\a}_{i} \|_1<1$) to the
  matrix $\tilde{\B}$ and to ${\Qc^\star\backslash\Qc}$ of size $k-g$.
  The atoms of $\tilde{\B}_{\Qc^\star\backslash\Qc}$ are of
    unit norm (actually, $\tilde{\B}_{\Qc^\star\backslash\Qc}$ is full
    rank) because $\A_{\Qcs\cup\Qc}$ is full
    rank~\cite[Corollary~3]{Soussen2013Joint}. 
\end{IEEEproof}\vspace{0.4cm}

The next lemma provides a useful upper bound on $\mu^{OLS}_l$ as a function of the coherence $\mu$ of the dictionary $\A$:
\begin{lemma} \label{prop:bound_mutilde}
If $\mu<1/l$, then
\begin{align}
  \mu^{OLS}_l\leq \frac{\mu}{1-l\mu}. \label{eq:bound_mutilde}
\end{align}
\end{lemma}

The proof of this result is reported to Appendix \ref{annex:SCOLS}.
The sufficient condition stated in Theorem \ref{th:mainth} for OLS$_\Qc$
then follows from the combination of Proposition \ref{lem:connect_OLS}
and Lemma \ref{prop:bound_mutilde}. Indeed, \eqref{eq:CBcondition_partial}
implies $\mu<1/(k+b-1)\leq 1/(g+b)$ since $2k-g+b-1=k+b-1+(k-g)> k+b-1\geq
  g+b$. Hence, the result follows by first applying
Lemma~\ref{prop:bound_mutilde} and \eqref{eq:CBcondition_partial}:
  \begin{align}
  \mu^{OLS}_{g+b}&\leq \frac{\mu}{1-(g+b)\mu} < \frac{1}{2k-2g-1},
\end{align}
and then Proposition~\ref{lem:connect_OLS}, which implies \eqref{eq:ERCOLS}. 
$\mu<1/(k+b-1)$ implies that the full rank assumption of Proposition~\ref{lem:connect_OLS} 
is met for any $\Qcs\cup\Qc$ of cardinality $k+b$~\cite[Lemma~2.3]{Tropp2004Greed}.

%\begin{theorem}[Partial Sufficient ERC for OLS]\label{th:mainth_ols}
%  Assume that, at the $j$th iteration, OLS has selected atoms in $\Qc
%  \subset \Qcs$, with $\vert \Qc\vert=j$, $\vert \Qcs\vert=k$. If
%  \begin{align}
%    \mu < \frac{1}{2k-j-1}, \label{eq:mainBound_copy}
%  \end{align}
%  then OLS exactly recovers the support of the sparse vector. 
%\end{theorem}
%
%\begin{IEEEproof}
%  \eqref{eq:mainBound_copy} implies that $\mu<1/j$ since
%  $2k-j-1=k+(k-j-1)\geq k\geq j$.  Apply
%  Lemma~\ref{prop:bound_mutilde}:
%  \begin{align*}
%  \mu^{OLS}_j&\leq \frac{\mu}{1-j\mu}{\color{red}<} \frac{1}{2k-2j-1}
%\end{align*}
%and then Lemma~\ref{lem:connect_OLS}: the ERC-OLS($\A,\Qc^\star,\Qc$)
%is met.
%\end{IEEEproof}

\section{Ordering properties and sufficiency of \eqref{eq:CBcondition_partial} for the success of $(P_{p,\Qc})$}\label{sec:proofsPp}

%\remCS{Ce debut de section doit etre repris avec des transitions pour
%  rendre la lecture plus confortable ! Actuellement, le lecteur doit
%  tourner les pages en arriere pour savoir a quoi se rapportent les
%  ths. 5,6 et 7 !}  
%\remCS{Il faut aussi rajouter une phrase pour expliquer que les
%resultats lies à Lp sont deduits d'un resultat qui connecte la
%condition L1 a la connexion OMP (th. 7), et c'est pour cette
%raison que ce resultat se trouve dans cette section (qui n'est 
%pas, autrement, dediee a OMP)}  
In this section, we elaborate on the proofs of Theorems \ref{th:mainthPp} (direct part),
\ref{th:nestingproperty} and \ref{th:nesting_ERCNSP}. These results have been gathered in this section since they are all related to some guarantees of success for $(P_{p,\Qc})$: Theorem \ref{th:mainthPp} shows that \eqref{eq:CBcondition_partial} is a sufficient and worst-case necessary condition for the success of $(P_{p,\Qc})$; Theorem \ref{th:nestingproperty} establishes an ordering property between the truncated NSPs for different values of $p\in[0,1]$; Theorem \ref{th:nesting_ERCNSP} emphasizes that the ERC-OMP \eqref{eq:ERCunif} is also a sufficient condition for the success of $(P_{1,\Qc})$ and in turn, of $(P_{p,\Qc})$ for $p<1$.

Theorems \ref{th:nestingproperty} and  \ref{th:nesting_ERCNSP} follow from some technical lemmas which are stated below and proved
  in Appendix \ref{app:ProofResultsPp}. The proof of the direct part of Theorem \ref{th:mainthPp} is a consequence of Theorems \ref{th:nestingproperty},  \ref{th:nesting_ERCNSP} and is discussed at the end of this section. 
The proof of the converse part of Theorem \ref{th:mainthPp} is reported to the next section. 

We first turn our attention to the proof of the NSP ordering stated in Theorem \ref{th:nestingproperty}. The result follows from the following lemma:
\begin{lemma} \label{lem:orderingNSP}  Assume $\spark(\A)>k+b$ and let $\forall \v\in\ker_0 (\A)$:
\begin{align}
\theta_p(k,g,b,\v) \triangleq  \max_{\vert \Qcs \vert = k} \max_{\substack{\vert \bar{\Qc}^\star\cap\Qc \vert = b\\ \vert \Qcs\cap\Qc \vert = g}} \biggl\{ \frac{\| \v_{\Qcs\backslash\Qc} \|_p^p}{\| \v_{\overline{\Qc\cup\Qcs}}\|_p^p} \biggr\}.
\label{eq:4dimensions}
\end{align}
Then, the following inequality holds for $0\leq q<p\leq 1$:
\begin{align}
\theta_q(k,g,b,\v) \leq \theta_p(k,g,b,\v). \label{lem:ordering1}
\end{align}
\end{lemma}
%
%The proof of the result is reported to Appendix \ref{app:ProofResultsPp}. 
Obviously, taking the supremum with respect to $\v\in\ker_0 (\A)$ of both sides in \eqref{lem:ordering1} leads to the result stated in Theorem \ref{th:nestingproperty}.
%\begin{IEEEproof}\textit{(Theorem  \ref{th:nestingproperty})}
%\end{IEEEproof}

Secondly, the inequality relating $\theta_{1}(k,g,b)$ to $\theta_{\mathrm{OMP}}(k,g,b)$ in Theorem \ref{th:nesting_ERCNSP} is a consequence of the next result:
\begin{lemma} \label{lem:orderingNSPERClem} If $\spark(\A)>k+b$, then 
\begin{align}
\frac{\| \v_{\Qcs\backslash\Qc}\|_1}{\|\v_{\overline{\Qcs\cup\Qc}}\|_1} \leq %\| \tA_{\Qcs\backslash\Qc}^\dag\tA_{\overline{\Qcs\cup\Qc}}\|_{1,1}
 \max_{i\notin{\Qcs}}\| \tA_{\Qcs\backslash\Qc}^\dag\tilde{\a}_i\|_1
 \label{eq:lem5}
 %,\nonumber\\[-0.2cm]
 %\nonumber
\end{align}
for any $\v\in\ker_0(\A)$ and $\Qcs$, $\Qc$ with $\vert \Qcs\vert=k$, $\vert \Qcs\cap\Qc\vert=g<k$, $\vert \bar{\Qc}^\star\cap\Qc\vert=b$.
\end{lemma}
%
%The proof is reported to Appendix \ref{app:ProofResultsPp}. 
Theorem \ref{th:nesting_ERCNSP} then follows by taking the supremum of both sides of \eqref{eq:lem5}
with respect to $\v\in\ker_0(\A)$ and $\Qcs$, $\Qc$ with $\vert \Qcs\vert=k$, $\vert \Qcs\cap\Qc\vert=g$ and $\vert \bar{\Qc}^\star\cap\Qc\vert=b$.

We are now ready to prove the sufficiency of \eqref{eq:CBcondition_partial} for $(P_{p,\Qc})$:\\[-0.2cm]
\begin{IEEEproof}\textit{(Direct part of Theorem  \ref{th:mainthPp})} On the one hand, let us first note that \eqref{eq:CBcondition_partial} is sufficient for the success of OMP$_\Qc$ for any $k$-sparse representation $\y=\A\xs$. Hence, \eqref{eq:CBcondition_partial} implies that $\theta_{\mathrm{OMP}}(k,g,b)<1$ since the latter condition is worst-necessary for the
success of OMP$_\Qc$ (Theorem~\ref{th:SoussenERC}). 
On the other hand, from \cite[Th.~2.4]{Tropp2004Greed}, we have that \eqref{eq:CBcondition_partial} is sufficient for $\spark(\A)>2k-g+b>k+b$. Applying successively Theorems \ref{th:nesting_ERCNSP} and \ref{th:nestingproperty}, we have
\begin{align}
\theta_{p}(k,g,b)<1 \qquad \forall p\in [0,1].
\end{align}
The result then follows from Theorem \ref{th:nestingpropertypartial}.
\end{IEEEproof}

\section{Worst-case necessity of \eqref{eq:CBcondition_partial}}\label{sec:NC}

%\remCS{Je te propose de rajouter deux sous-section pour que ce soit plus clair qu'a la fin c'est le cas particulier b=0 qui est traite}

\subsection{General case $b\geq 0$}
Cai\&Wang recently showed in \cite[Th. 3.1]{Cai2010Stable} that
there exist dictionaries $\A$ with $\mu = \frac{1}{2k-1}$ and 
%linear
%combinations $\y$ of $k$ columns of $\A$ such that $\y$ has 
a vector $\y\in\spa (\A)$ having \emph{two}
disjoint $k$-sparse representations in $\A$. In other words, if $\mu <
\frac{1}{2k-1}$ is not satisfied, there exist instances of
dictionaries such that \emph{no} algorithm can univocally recover some
$k$-sparse representations. In the context of Oxx, their result can be
rephrased as the following worst-case necessary condition: there
exists a dictionary $\A$ with $\mu = \frac{1}{2k-1}$ and a support
$\Qcs$, with $\vert \Qcs \vert =k$, such that Oxx selects a wrong atom
at the first iteration.

In this section, we show that  \eqref{eq:CBcondition_partial} is worst-case necessary for $(P_{p,\Qc})$ and Oxx$_\Qc$ in the sense defined in 
Theorems \ref{th:mainth} and \ref{th:mainthPp}, respectively. 
To prove the result for $(P_{p,\Qc})$, we will construct a dictionary $\A$ satisfying $\mu=\frac{1}{2k-g+b-1}$ and such that
\begin{align}
\theta_p(k,g,b)=1\qquad \forall p\in[0,1]. \label{eq:NSPAWC}
\end{align}
The result then immediately follows from Theorem \ref{th:nestingpropertypartial}. 
 Invoking Theorem \ref{th:nesting_ERCNSP} and the converse part
  of Theorem \ref{th:SoussenERC}, \eqref{eq:NSPAWC} also leads to the
  result for OMP$_\Qc$: in particular,
  $\theta_{\mathrm{OMP}}(k,g,b)\geq 1$. On the other hand, since
  $\theta_{\mathrm{OLS}}(k,g,b)$ and $\theta_1(k,g,b)$ do not enjoy a
  nesting property similar to \eqref{eq:orderingL1OMP}, specific
  arguments need to be derived to prove the worst-case necessity of
  \eqref{eq:CBcondition_partial} for OLS$_\Qc$. 
 Regarding OLS$_\Qc$ (and actually, also OMP$_\Qc$), we will show using the same dictionary as for $(P_{p,\Qc})$, that there exists a $k$-term representation $\y=\A\xs$  satisfying the hypotheses of Theorem \ref{th:mainth} and such that Oxx$_\Qc$ selects a wrong atom at the first iteration. 
%Finally, a demonstration of Lemma \ref{lem:Oxxwc2} is provided at the end of this section. 
The proofs for  Oxx$_\Qc$ and $(P_{p,\Qc})$ use a dictionary construction similar to Cai\&Wang's in \cite{Cai2010Stable}.  \\

%We adopt an approach similar to Cai\&Wang's in \cite{Cai2010Stable} for the dictionary construction. 
Let $\G \in\mathbb{R}^{(2k-g+b) \times (2k-g+b)}$ be the matrix with ones on the diagonal and 
$-\mu\triangleq-\frac{1}{2k-g+b-1}$ elsewhere. $\G$ will play the
  role of the Gram matrix $\G=\A^T\A$. We will exploit the eigenvalue
decomposition of $\G$ to construct the dictionary
$\A\in\mathbb{R}^{(2k-g+b-1)\times (2k-g+b)}$ with the desired properties.
Since $\G$ is symmetric, it can be expressed as
\begin{align}
\G = \U \Diag \U^T,
\end{align}
where $\U$ (resp. $\Diag$) is the unitary matrix whose
  columns are the eigenvectors (resp. the diagonal matrix of
eigenvalues) of $\G$. It is easy to check
(see Example \ref{exple:RICmu1}) that $\G$ has only two
distinct eigenvalues: 
$1+\mu$ with multiplicity $2k-g+b-1$
and $0$ with multiplicity one; moreover, the eigenvector associated to
the null eigenvalue is equal to $\mathbf{1}_{2k-g+b}$. The eigenvalues
are sorted in the decreasing order so that $0$ appears in the lower
right corner of $\Lambda$.\\

We  define $\A\in\mathbb{R}^{(2k-g+b-1)\times (2k-g+b)}$ as
\begin{align}
\A = \Upsilon \U^T, \label{eq:defDico}
\end{align}
where $\Upsilon \in \mathbb{R}^{(2k-g+b-1)\times (2k-g+b)}$ is such that
\begin{align}
\Upsilon(i,j) = \left\{
\begin{array}{cl}
\sqrt{1+\mu} & \mbox{if $i=j$,}\\
0 & \mbox{otherwise.}
\end{array}
\right.
\end{align}
Note that $\Upsilon^T  \Upsilon=\Lambda$. Hence, $\A$ satisfies the statement (i) in the converse part of Theorems \ref{th:mainth} and \ref{th:mainthPp} since
\begin{align}
\A^T \A = \U \Upsilon^T  \Upsilon \U^T = \U \Diag \U^T = \G,
\end{align}
and therefore 
\begin{align}
\langle \a_i, \a_j \rangle = -\mu \quad \forall i \neq j.
\label{eq:prod_scal_atomes}
\end{align}
Since $\G=\A^T\A$, we have $\G\x=\mathbf{0}_{2k-g+b}$ if and only if
$\A\x=\mathbf{0}_{2k-g+b}$. Moreover, since $\G$ has \emph{one single}
zero eigenvalue with eigenvector $\mathbf{1}_{2k-g+b}$, the
  null-space of $\A$ is the one-dimensional space spanned by
  $\mathbf{1}_{2k-g+b}$. Therefore, any $l<2k-g+b$ columns of $\A$ are
linearly independent, \ie $\spark(\A)=2k-g+b>k+b$.

Taking these observations into account, it easily follows that \eqref{eq:NSPAWC} holds since
\begin{align}
\| \v_{\Qcs\backslash\Qc}\|^p_p&=\| \v_{\overline{\Qcs\cup\Qc}}\|^p_p=k-g,
\end{align}
for $\v= \mathbf{1}_{2k-g+b}\in \ker_0(\A)$, and $\forall\,\Qcs$,$\Qc$
with $\vert \Qcs\vert=k$, $\vert \Qcs\cap\Qc\vert=g$ and $\vert
\bar{\Qc}^\star\cap\Qc\vert=b$. %This proves the worst-case necessity of \eqref{eq:CBcondition_partial} for $(P_{p,\Qc})$. 
This proves the necessary part of Theorem~\ref{th:mainthPp}.

%\begin{theorem}[\eqref{eq:CBcondition_partial} is a worst-case necessary condition for Oxx]\label{th:NC}
%There exists a dictionary $\A$ with $\mu=\frac{1}{2k-l-1}$, a support $\Qcs$ with $\vert \Qcs \vert=k$ and $\y\in\spa(\A_\Qcs)$, such that 
%Oxx with $\y$ as input selects $l$ atoms in $\Qcs$ during the first $l$ iterations and a wrong atom at the $(l+1)$th iteration.
%%\begin{itemize}
%%\item[\emph{i)}] Oxx selects $l$ atoms in $\Qcs$ during the first $l$ iterations,
%%\item[\emph{ii)}] Oxx selects a wrong atom at the $(l+1)$th iteration. 
%%\end{itemize}
%\end{theorem}

%Let us note that, by invoking Theorem \ref{th:nesting_ERCNSP} and the converse part of Theorem \ref{th:SoussenERC}, \eqref{eq:NSPAWC} and $\spark(\A)>k+b$ also lead to the result for OMP$_\Qc$. On the other hand, since $\theta_{\mathrm{OLS}}(k,g,b)$ and $\theta_1(k,g,b)$ do not enjoy a nesting property similar to \eqref{eq:orderingL1OMP},  specific arguments need to be derived to prove the worst-case necessity of \eqref{eq:CBcondition_partial} for OLS.  For the sake of generality, we develop our arguments for both OMP and OLS hereafter.  

We now address the case of OLS. Although the OMP
  necessity result is already obtained from the $(P_{p,\Qc})$
  necessity result, the construction related to OLS
  is also valid for OMP. For the sake of generality,
we develop our arguments for both OMP and OLS hereafter. 
 We first need the following technical lemma whose proof is reported to Appendix \ref{app:NC}:
\begin{lemma}\label{lem:Ctilde}
  Let $\A$ be defined as in \eqref{eq:defDico}.
    %Then, for any subset $\Qc$ with $\vert \Qc \vert=l$, we have $\forall i\notin\Qc,\,\tc_i^{\Qc}\ne\mathbf{0}_m$.} Moreover, 
    Then, for any subset $\Qc$ with $\vert \Qc \vert=g+b$, there
    exists a vector $\tilde{\y}$ having two $(k-g)$-term representations
    with disjoint supports in the projected dictionary
    $\tC_{\backslash\Qc}\triangleq \tC_{\{1,\ldots,
      2k-g+b\}\backslash\Qc} \in \mathbb{R}^{(2k-g+b-1)\times(2k-2g)}.$ 
\end{lemma}

We are now ready to prove the worst-case necessity of \eqref{eq:CBcondition_partial} for Oxx$_\Qc$:\\

\begin{IEEEproof}\emph{(Converse part of Theorem \ref{th:mainth})}
  We show that there exists a $k$-sparse representation $\y$ such that
  Oxx$_\Qc$ selects a wrong at the first iteration with the dictionary $\A$ defined in
    \eqref{eq:defDico}. Our construction of such $\y$ is as follows.
  Let $\Qc$ be a subset of cardinality $g+b$, arbitrarily chosen (say,
  the first $g+b$ atoms of the dictionary). We consider the following decomposition  $\Qc=\Qc_g\cup\Qc_b$ where $\Qc_g$ and $\Qc_b$ are the subsets collecting respectively the good and the bad atoms in $\Qc$, with $\Qc_g\cap\Qc_b=\emptyset$. Let $\tilde{\y}_2$ be a vector having two 
 $(k-g)$-term representations in the projected dictionary
    $\tC_{\backslash\Qc}$. We note that such a vector $\tilde{\y}_2$ exists by virtue of Lemma~\ref{lem:Ctilde}. We will denote the respective supports of the two representations of $\tilde{\y}_2$ by  $\Qc_1$ and $\Qc_2$ with $\Qc_1\cap\Qc_2=\emptyset$. Hence,
\begin{align}
\tilde{\y}_2=\tilde{\C}_{\Qc_1} \x_{\Qc_1}=\tilde{\C}_{\Qc_2} \x_{\Qc_2},
\end{align}
for some vectors $\x_{\Qc_1}$ and $\x_{\Qc_2}$.
We then define the $k$-sparse representation
\begin{align}
\y \triangleq \y_1 +\y_2,
\end{align}
where $\y_1=\A_{\Qc_g}\oneb_{\vert\Qc_g\vert}$ and $\y_2=\A_{\Qc_i} \x_{\Qc_i}\in \spa(\A_{\Qc_i})$ with $i=1$ or $i=2$. The specific value of $i$ will be determined hereafter so that a failure situation occurs.\\
    The
    selection rule~\eqref{eq:atomselection2} indicates that the atom
    $\ta_j$ selected by Oxx$_\Qc$ at the first iteration satisfies:
    \begin{align}
      j\in \ama_i  \vert \stdscal{\tc_i,\proj \y} \vert = \ama_i  \vert \stdscal{\tc_i,\tilde{\y}_2 } \vert,\label{eq:Oxxselecstep2}
    \end{align}
    since $\proj \y = \proj \y_2 = \tilde{\y}_2$.
      Now, we set $\Qcs$ in such a way that $j\notin\Qcs$:
    \begin{align}
      \Qcs=\left\{
        \begin{array}{cl}
          \Qc_g\cup\Qc_1 & \mbox{if $j\in\Qc_2$},\\
          \Qc_g\cup\Qc_2 & \mbox{if $j\in\Qc_1$.}
        \end{array}
      \right .
      \label{eq:Qcs}
    \end{align}
      To complete the proof, it is easy to check that
      $\y=\y_1+ \y_2\in\spansub{\A_{\Qcs}}$: indeed,
      $\y_1\in\spansub{\A_{\Qc_g}}\subset \spansub{\A_{\Qcs}}$ and
      $\y_2\in\spansub{\A_{\Qcs\backslash\Qc}}
      \subset\spansub{\A_{\Qcs}}$. 
\end{IEEEproof}\vspace{0.5cm}

\subsection{Special case $b=0$}
We now turn our attention to the proof of Lemma \ref{lem:Oxxwc2}, which is related to the standard version of
  Oxx, initialized with the empty support. We first need to define the concept of ``reachability" of a subset $\Qc$:%the following technical lemma which provides sufficient conditions under which  OMP selects a certain atom at a given iteration: 
\begin{defi}\label{def:reachability}
A subset $\Qc$ is said to be reachable by Oxx if there exists $\y \in \spa (\A_\Qc)$ such that Oxx with $\y$ as input selects atoms in $\Qc$ during the first $\vert \Qc \vert$ iterations. 
\end{defi}

The concept of reachability was first introduced in
\cite{Soussen2013Joint}. The authors showed that any subset $\Qc$
with $\vert \Qc \vert\leq \spark(\A)-2$ is reachable by OLS, see
\cite[Lemma 3]{Soussen2013Joint}. On the other hand, they emphasized
that there exist dictionaries for which some subsets $\Qc$ can never
be reached by OMP, see
\cite[Example 1]{Soussen2013Joint}. This scenario does however not
occur for the dictionary defined in \eqref{eq:defDico} as stated in
the next lemma:
\begin{lemma}\label{lem:SCOMP_NC}
Let $\A$ be defined as in \eqref{eq:defDico} with $b=0$. Then any subset $\Qc$ with $\vert \Qc \vert=g$ is reachable by Oxx. 
\end{lemma}

%\begin{lemma} \label{lem:SCOMP_NC} Let $\A\in\mathbb{R}^{(2k-l-1)\times (2k-l)}$ be defined as in \eqref{eq:defDico}. Assume that OMP with input $\y=\A_\Qcs \x_\Qcs$ has selected atoms in $\Qc\subset\Qcs$ during the first $l$ iterations. If
%for some $j\in \Qcs\backslash\Qc$, the following conditions hold:
%\begin{align}
%& sign(x_j) = - sign(\sum_{i\in\Qcs\backslash\Qc} x_i), \label{eq:condl3_1}\\
%& \vert x_j \vert > \vert x_i \vert\quad\mbox{$\forall\, i\in \Qcs\backslash(\Qc\cup\{j\}) $}, \label{eq:condl3_2}
%\end{align}
%then OMP selects atom $j$ at the next iteration. 
%\end{lemma}
The proof of this result is reported to Appendix \ref{app:NC}. 
We are now ready to prove Lemma \ref{lem:Oxxwc2}: \vspace{0.2cm}

\begin{IEEEproof}\emph{(Lemma \ref{lem:Oxxwc2})}
Consider the dictionary $\A$ defined in \eqref{eq:defDico}
    with $b=0$. Let $\Qc$ be a subset of cardinality $g$, arbitrarily
    chosen (say, the first $g$ atoms of the dictionary). We will
    exhibit a subset $\Qcs\supset\Qc$ for which the result of
    Lemma \ref{lem:Oxxwc2} holds.

  We first apply Lemma~\ref{lem:SCOMP_NC}: there exists an
    input $\y_1\in\spa(\A_\Qc)$ for which Oxx selects all atoms in
    $\Qc$ during the first $g$ iterations. Then, we apply
    Lemma~\ref{lem:Ctilde}: there exists a vector $\tilde{\y}_2$ having two
    $(k-g)$-term representations in the projected dictionary
    $\tC_{\backslash\Qc}$. We will denote their respective supports by
    $\Qc_1$ and $\Qc_2$ with $\Qc_1\cap\Qc_2=\emptyset$. We then define $\y_2$ as in the proof of the converse of Theorem \ref{th:mainth}. 

  By virtue of \cite[Lemma~15]{Soussen2013Joint}, Oxx with
    $\y=\y_1+\varepsilon \y_2$ as input selects the same atoms (\ie
    $\Qc$) as with $\y_1$ as input during the first $g$ iterations as
    long as $\varepsilon>0$ is sufficiently small. Moreover, defining $\Qcs$ as in \eqref{eq:Qcs} and applying the same reasoning as in the proof of the converse part of Theorem \ref{th:mainth}, we have that $\y\in\spansub{\A_{\Qcs}}$ and is such that Oxx selects a bad atom at iteration $g+1$.   
%    the
%    selection rule~\eqref{eq:atomselection2} indicates that the atom
%    $\ta_j$ selected at iteration $l+1$ satisfies:
%    \begin{align}
%      j\in \ama_i  \vert \stdscal{\tc_i,\proj \y} \vert = \ama_i  \vert \stdscal{\tc_i,\y_2} \vert,\label{eq:Oxxselecstep2}
%    \end{align}
%    since $\proj \y = \varepsilon \proj \y_2 = \varepsilon \y_2$.
%      Now, we set $\Qcs$ in such a way that $j\notin\Qcs$:
%    % 
%    \begin{align}
%      \Qcs=\left\{
%        \begin{array}{cl}
%          \Qc\cup\Qc_1 & \mbox{if $j\in\Qc_2$},\\
%          \Qc\cup\Qc_2 & \mbox{if $j\in\Qc_1$.}
%        \end{array}
%      \right .
%      \label{eq:Qcs}
%    \end{align}
%    % 
%    To complete the proof, it is easy to check that
%      $\y=\y_1+\varepsilon \y_2\in\spansub{\A_{\Qcs}}$ because
%      $\y_1\in\spansub{\A_{\Qc}}$ and
%      $\y_2\in\spansub{\tC_{\Qcs\backslash\Qc}}=\spansub{\tA_{\Qcs\backslash\Qc}}
%      \subset\A_{\Qcs}$.
 \end{IEEEproof}

 \section{Quasi-tightness of \eqref{eq:RICCOMPQ} for OMP$_\Qc$}
 \label{sec:quasitighnessRIC} In this section, we provide an instance
 of dictionary such that $\delta_{k+b+1}=1/\sqrt{k-g}$ and Oxx$_\Qc$
 fails at the first iteration.
 Our dictionary construction is along the same line as \cite[Theorem 3.2]{Mo2012Remark}. \\

\begin{IEEEproof} \textit{(Lemma \ref{lem:quasi-tighnessRIC})}
We first consider the case $g\leq k-2$. Let us define $\A\in\R^{(k+b+1)\times (k+b+1)}$ as
\begin{align}\label{eq:defA}
\A &=
\left(
\begin{array}{cc}
\I_{g+b} & \mathbf{0}_{(g+b)\times (k-g+1)}\\
\mathbf{0}_{(k-g+1) \times (g+b)} & \M
\end{array}
\right)
\end{align}
where
\begin{align}
\M \triangleq 
\left(
\begin{array}{cccc}
 & & & \frac{1}{k-g}\\
 & \I_{k-g} & & \vdots\\
 & & & \frac{1}{k-g}\\
0 & \ldots & 0 & \sqrt{\frac{k-g-1}{k-g}}
\end{array}
\right)
\end{align}

On the one hand, it can be seen that the eigenvalues of the Gram matrix $\G=\A^T \A$ are $\lambda=1$ with multiplicity 
$k+b-1$ and
 $\lambda=1\pm \frac{1}{\sqrt{k-g}}$ with multiplicity 1. Hence, $\delta_{k+b+1}=\frac{1}{\sqrt{k-g}}$. 

On the other hand, there exist $\Qcs$ and $\Qc$ satisfying the
hypotheses of Lemma \ref{lem:quasi-tighnessRIC} and such that
Oxx$_\Qc$ fails at the first iteration for some representation
  $\y=\A\xs$ indexed by $\Qcs$.  Let us set $\Qc=\{1,\ldots,g+b\}$,
$\Qcs=\{b+1,\ldots,k+b\}$ in such a way that there is only 
one wrong atom outside of $\Qc\cup\Qcs$, namely the last atom. We set
\begin{align}
x^\star_i &=\left\{
\begin{array}{ll}
1 & \mbox{if}\; i\in \Qcs\\
0 & \mbox{otherwise}.
\end{array} \right.
\end{align}
With this particular choice, we have $\y=\A_{\Qcs}\mathbf{1}_k$ and:
\begin{align}
\tC_{\{1,\ldots, k+b+1\}\backslash\Qc}&=
\A_{\{1,\ldots, k+b+1\}\backslash\Qc} =
\left(
\begin{array}{c}
\mathbf{0}_{(g+b)\times (k-g+1)}\\
\M
\end{array}
\right)\\
\r^\Qc&=\tC_{\Qcs\backslash\Qc}\mathbf{1}_{k-g}=
\left(
\begin{array}{c}
\mathbf{0}_{g+b}\\
\mathbf{1}_{k-g}\\
0
\end{array}
\right)
\end{align}
and therefore,
\begin{align}
\langle \tc_i, \r^\Qc\rangle = 1 \quad\forall i\geq g+b+1.
\end{align}
Since $k+b+1 \notin \Qcs$, a failure situation as in \eqref{eq:equalitySR} occurs.  

%Finally, the case where $g=k-1$ can be handled similarly by defining $\A$ as in \eqref{eq:defA} with 
The special case $g=k-1$ leads to the degenerate
  situation $\delta_{k+b+1}=1$ in Lemma~\ref{lem:quasi-tighnessRIC}.
  This case is handled by proposing a dictionary having two identical
  columns. We define $\A$ as in \eqref{eq:defA} with 
\begin{align}
\M=\left (
\begin{array}{cc}
 1& 1\\
 0& 0
\end{array}
 \right)\in \mathbb{R}^{2\times 2}.
\end{align}
We have obviously $\delta_{k+b+1}=1$  since that the dictionary has two
identical columns. Oxx$_\Qc$ then trivially fails with $\y$, $\Qcs$ and $\Qc$ defined as
above.
%\begin{align}
%\M=(1\quad 1)\in \mathbb{R}^{1\times 2}.
%\end{align}
%
%In this case, $\delta_{k+b+1}=1$ since that the dictionary as two identical columns and Oxx$_\Qc$ then trivially fails.

\end{IEEEproof}

\section{Conclusions}

We derived a new sufficient and worst-case necessary condition, $\mu<\frac{1}{2k-g+b-1}$, for the success of OMP, OLS and some procedures based on $\ell_p$ relaxation. Our result both applies to the context of sparse representations with support side information, and to the analysis of greedy algorithms at intermediate iterations. Our condition relaxes the well-known coherence-based result $\mu<\frac{1}{2k-1}$ derived in the non-informed setup by several authors, see \eg \cite{Tropp2004Greed, Fuchs2004Sparse, Gribonval2003Sparse}. Moreover, it  is shown to be complementary with some similar conditions based on restricted isometry constants \cite{Karahanoglu2012Theoretical, Friedlander2012Recovering}.

We also carried out a fine analysis of some relations existing between conditions of success for OMP/OLS and $\ell_p$-relaxed procedures in the informed setup. We showed that the truncated NSP, characterizing the success of $\ell_p$-relaxed procedures in the informed setup, enjoys some ordering property. Moreover, we established a direct implication between the ERC-OMP derived in \cite{Soussen2013Joint} and the truncated NSP for the informed $\ell_1$-relaxed problem.

\appendices
\section{Proof of the results of section \ref{sec:SCOMP}}

This section contains the proofs of Lemmas~\ref{UBERC} and
\ref{prop:linkRICmu} together with some useful technical lemmas.
%We first need to 
%define the coherence\footnote{The
%  standard definition of the coherence usually assumes that the
%  columns of the dictionary are normalized. We thus use the term
%  "coherence" with a slight abuse of language here since the columns
%  of $\proj \A$ are not necessarily normalized to 1.} of the projected
%dictionary as follows:
% and drive a link between the latter and the upper/lower bounds defined in Definition \ref{def:GRIC}:
%
%\begin{defi}
%\begin{align}
%\mu^{OMP}_l= \max_{\vert \Qc \vert=l}\max_{i \neq j}  \vert \langle \ta_i, \ta_j  \rangle\vert. 
%\end{align}
%\end{defi}
%When $l=0$, one recovers the standard definition of the coherence of the dictionary.
% The next lemma relates the coherence of the projected dictionary to the isometry bounds of the projected dictionary $\ud_{2,l}$ and $\ld_{2,l}$:
\begin{lemma}\label{len:lemma1} Assume $\A$ satisfies the  P-RIP($\ld_{2,l}$,$\ud_{2,l}$) and let
\begin{align}
\mu^{OMP}_l\triangleq \max_{\vert \Qc \vert=l}\max_{i \neq j}  \vert \langle \ta_i^\Qc, \ta_j^\Qc  \rangle\vert. 
\end{align}
Then, we have
\begin{align}
%\mu^{OMP}_l \leq \frac{\ub(2,l)-\lb(2,l)}{2}.
\mu^{OMP}_l \leq \frac{\ud_{2,l}+\ld_{2,l}}{2}.
\end{align}
\end{lemma}
\begin{IEEEproof}
By definition of $\ud_{2,l}$ and $\ld_{2,l}$ we must have for all $\Qc, \Qc'$ with $\vert \Qc \vert=l$, $\vert \Qc' \vert=2$ and $\Qc'\cap\Qc=\emptyset$:
\begin{align}
1+\ud_{2,l}\geq \lambda_{max}(\tilde{\A}_{\Qc'}^T \tilde{\A}_{\Qc'}),\label{eq:lbub}\\
1-\ld_{2,l} \leq \lambda_{min}(\tilde{\A}_{\Qc'}^T \tilde{\A}_{\Qc'}),\label{eq:ublb}
\end{align}
where $\lambda_{max}(\M)$ (resp. $\lambda_{min}(\M)$) denotes the
largest (resp. smallest) eigenvalue of $\M$ and the tilde notation is relative to subset
$\Qc$. Moreover, if
${\Qc'}=\{i,j\}$, it is easy to check that the eigenvalues of
$\tilde{\A}_{\Qc'}^T \tilde{\A}_{\Qc'}$ can be expressed as
\begin{align}
\lambda_{}(\tilde{\A}_{\Qc'}^T \tilde{\A}_{\Qc'})=\frac{\| \tilde{\a}_i\|^2+\| \tilde{\a}_j\|^2 \pm \Delta}{2},\nonumber
%\lambda_{min}(\tilde{\A}_{\Qc'}^T \tilde{\A}_{\Qc'})=\frac{\| \tilde{\a}_i\|^2+\| \tilde{\a}_j\|^2 - \Delta}{2},\nonumber
\end{align}
where
\begin{align}
\Delta &=\sqrt{(\| \tilde{\a}_i\|^2+\| \tilde{\a}_j\|^2)^2+4(
    \langle \tilde{\a}_i, \tilde{\a}_j \rangle^2-\| \tilde{\a}_i\|^2
    \: \| \tilde{\a}_j\|^2) }  \\
&=\sqrt{(\| \tilde{\a}_i\|^2-\| \tilde{\a}_j\|^2)^2+4 \langle \tilde{\a}_i, \tilde{\a}_j \rangle^2 }. 
\end{align}
Hence
\begin{align}
\lambda_{max}(\tilde{\A}_{\Qc'}^T \tilde{\A}_{\Qc'})-\lambda_{min}(\tilde{\A}_{\Qc'}^T \tilde{\A}_{\Qc'})
&= \Delta \geq 2 \vert \langle \tilde{\a}_i, \tilde{\a}_j \rangle \vert. \nonumber
\end{align}
Using \eqref{eq:lbub}-\eqref{eq:ublb}, we thus obtain $\forall i, j \notin \Qc$:
\begin{align}
\ud_{2,l}+\ld_{2,l} \geq 2 \vert \langle \tilde{\a}_i, \tilde{\a}_j \rangle \vert.\label{eq:ineqinterqqq}
\end{align}
Now, this inequality also holds if $i\in \Qc$ or $j\in \Qc$ since the
right hand-side of \eqref{eq:ineqinterqqq} is then equal to zero.  The
result then follows from the definition of $\mu_l^{OMP}$.
\end{IEEEproof}

\begin{lemma} \label{len:lemma2} Let $\vert \Qc \vert$=l and $\Qc' \cap \Qc'' =\emptyset$, then $\forall \u \in \mathbb{R}^{\vert\Qc''\vert}$,
\begin{align}
\| \tA_{\Qc'}^T  \tA_{\Qc''} \u  \|\leq  \mu^{OMP}_l \sqrt{\vert {\Qc'} \vert \vert {\Qc''} \vert} \, \|\u\|.
\end{align}
\end{lemma}

 \begin{IEEEproof}
 We have
 \begin{align}
\| \tilde{\A}_{\Qc'}^T \tilde{\A}_{\Qc''} \u  \| 
&= \sqrt{\sum_{i\in {\Qc'}} \langle \tilde{\a}_i, \tilde{\A}_{\Qc''} \u \rangle^2}\\
&=\sqrt{\sum_{i\in {\Qc'}} 
\bigl (\sum_{j\in {\Qc''}} u_j\, \langle \tilde{\a}_i, \tilde{\a}_j \rangle
\bigr)^2}\\
&\leq \sqrt{\sum_{i\in {\Qc'}} \bigl(\sum_{j\in {\Qc''}} \vert u_j\vert \, \vert\langle \tilde{\a}_i, \tilde{\a}_j \rangle\vert\bigr)^2}\\
&\leq  \mu^{OMP}_l \sqrt{\vert {\Qc'} \vert}\, \| \u \|_1\\
&\leq  \mu^{OMP}_l \sqrt{\vert {\Qc'} \vert \vert {\Qc''} \vert}\, \| \u \|.
\end{align}
\end{IEEEproof}

Using Lemmas~\ref{len:lemma1} and~\ref{len:lemma2}, we can now prove Lemmas \ref{UBERC} and \ref{prop:linkRICmu}:\vspace{0.2cm}

\begin{IEEEproof}
\emph{(Lemma \ref{UBERC})} $\forall\, i\notin \Qcs$, the following inequalities 
%\remCS{ci-dessous, la notation $\|.\|_2$ a remplacer par $\|.\|$ ou alors remplacer partout $\|.\|$ par $\|.\|_2$} 
hold:
\begin{align}
\| \tA_{\Qc^\star \backslash \Qc}^\dag \tilde{\a}_{i} \|_1 
&\leq \sqrt{k-g} \,\| \tA_{\Qc^\star \backslash \Qc}^\dag \tilde{\a}_{i} \|,\\
&\leq \frac{\sqrt{k-g}}{1-\ld_{k-g,g+b}} \| \tA_{\Qc^\star \backslash \Qc}^T \tilde{\a}_{i} \|,\\
&\leq \frac{k-g}{1-\ld_{k-g,g+b}} \mu^{OMP}_{g+b},\\
&\leq \frac{k-g}{1-\ld_{k-g,g+b}} \frac{\ud_{2,g+b}+\ld_{2,g+b}}{2},
\end{align}
where the first inequality follows from the equivalence of norms; the second from RIC properties (see \cite[Prop. 3.1]{Needel_ACMels08}); the third from Lemma \ref{len:lemma2} and the fourth from Lemma \ref{len:lemma1}. 
\end{IEEEproof}\vspace{0.2cm}

\begin{IEEEproof}
\emph{(Lemma \ref{prop:linkRICmu})} 
First, notice that $\A$ satisfies the P-RIP($\ld_{q,0}$,$\ud_{q,0}$) $\forall \, q$ with
\begin{align}
\ud_{q,0}=\ld_{q,0}&= (q-1)\mu,
%\ud_{q,0}&= (q-1)\mu,
\end{align}
%\begin{align}
%1-(q-1)\mu \leq \lb(q,0) \leq \ub(q,0)\leq 1+(q+1)\mu,\quad \forall q
%\end{align}
see \eg \cite[Lemma 2.3]{Tropp2004Greed}. 
Hence, \eqref{eq:relRICmu1} is a consequence of the following inequalities:
\begin{align}
 \| \proj \A_{\Qc '} \xQp \|^2 \leq  \| \AQp \xQp \|^2 \leq (1+\ud_{q,0}) \|\xQp \|^2.
\end{align}
Lower bound \eqref{eq:relRICmu2} is derived by noticing that
\begin{align}
\| \proj \AQp \xQp \|^2 &= \| \AQp \xQp \|^2 -\| \mathbf{P}_\Qc\AQp \xQp \|^2,
\end{align}
and 
\begin{align}
\| \AQp \xQp \|^2&\geq (1-\ld_{q,0}) \|\xQp \|^2,\\
\| \mathbf{P}_\Qc\AQp \xQp \|^2
&=\|(\A_\Qc^\dag)^T \A_\Qc^T \A_{\Qc'} \x_{\Qc'}\|^2\\
&\leq \frac{\|\A_\Qc^T \A_{\Qc'} \x_{\Qc'}\|^2}{1-\ld_{l,0}}\label{eq:ineqPRIP},\\
&\leq \frac{\mu^2 l q \, \| \x_{\Qc'}\|^2}{1-\ld_{l,0}},\label{eq:conlemma2}
\end{align}
where inequality \eqref{eq:ineqPRIP} follows from standard
  relationships between the RIC properties of $\A$ and transforms of
  $\A$,  and $1-\ld_{l,0}\geq 0$ is a consequence of hypothesis
$\mu<1/(l-1)$~\cite[Lemma~2.3]{Tropp2004Greed}; \eqref{eq:conlemma2}
is a consequence of Lemma \ref{len:lemma2}.
\end{IEEEproof}

\section{Proof of the results of section \ref{sec:SCOLS}}\label{annex:SCOLS}

\begin{IEEEproof} \emph{(Lemma \ref{prop:bound_mutilde})}
The proof is recursive. Obviously, the result holds for 
$l=0$ since $\mu^{OLS}_0=\mu$.

Let $\Qc$ with $\vert\Qc\vert=l\geq1$ and consider $\Rc$ such that
$\Qc=\Rc\cup{\{i\}}$ with $\vert\Rc\vert=l-1$. According
to~\cite[Lemma~5]{Soussen2013Joint},
if $j\notin\Qc$, we have the orthogonal decomposition
\begin{align}
  \tilde{\b}_{j}^{\Rc} = \eta_{j}\tilde{\b}_{j}^{\Qc}+
  \langle \tilde{\b}_{j}^{\Rc} \,,\, \tilde{\b}_{i}^{\Rc} \rangle
  \,\tilde{\b}_{i}^{\Rc}.
  \label{eq:orthog_decomp}
\end{align}
 Moreover, assumption $\mu<1/l$ implies that
  $\A_{\Qc\cup\{j\}}$, $\A_{\Rc\cup\{j\}}$ and $\A_{\Rc\cup\{i\}}$ are
  full column rank as families of at most $l+1$
  atoms~\cite[Lemma~2.3]{Tropp2004Greed} which in turn implies that
  $\tilde{\a}_{j}^{\Qc}$, $\tilde{\a}_{j}^{\Rc}$ and
  $\tilde{\a}_{i}^{\Rc}$ are
  nonzero~\cite[Cor.~3]{Soussen2013Joint}. Therefore,
  $\|\tilde{\b}_{j}^{\Qc}\|$, $\|\tilde{\b}_{j}^{\Rc}\|$ and
  $\|\tilde{\b}_{i}^{\Rc}\|$ are all of unit norm, and 
then~\eqref{eq:orthog_decomp} yields $\eta_j=\pm \sqrt{1-\langle
  \tilde{\b}_{j}^{\Rc} \,,\, \tilde{\b}_{i}^{\Rc} \rangle^2}$.
If $j$ and $j'\notin\Qc$, it follows that 
\begin{align}
  \langle \tilde{\b}_{j}^{\Qc}\,,\,\tilde{\b}_{j'}^{\Qc}\rangle&=
  \frac{
  \langle \tilde{\b}_{j}^{\Rc}\,,\,\tilde{\b}_{j'}^{\Rc} \rangle -
  \langle \tilde{\b}_{j}^{\Rc} \,,\, \tilde{\b}_{i}^{\Rc} \rangle
  \langle \tilde{\b}_{j'}^{\Rc} \,,\, \tilde{\b}_{i}^{\Rc} \rangle}{
  \eta_{j}\eta_{j'}}.
\end{align}
Majorizing the inner products 
on the right-hand side by $\mu^{OLS}_{l-1}$
and using~\eqref{eq:bound_mutilde}, we get:
\begin{align}
  \vert\langle\tilde{\b}_{j}^{\Qc}\,,\,\tilde{\b}_{j'}^{\Qc}\rangle
  \vert&\leq \frac{\mu^{OLS}_{l-1}+(\mu^{OLS}_{l-1})^2}{
  1-(\mu^{OLS}_{l-1})^2}\\
&= \frac{\mu^{OLS}_{l-1}}{1-\mu^{OLS}_{l-1}}\\
&\leq \frac{\mu}{1-(l-1)\mu-\mu}=\frac{\mu}{1-l\mu}
\end{align}
leading to~\eqref{eq:bound_mutilde}.
\end{IEEEproof}

\section{Proof of the results of section \ref{sec:proofsPp}} \label{app:ProofResultsPp}
Before proceeding to the proofs of Lemmas \ref{lem:orderingNSP} and \ref{lem:orderingNSPERClem}, we emphasize that  $\spark(\A)>k+b$ and $\v\in\ker_0{(\A)}$ are sufficient conditions for $\v_{\overline{\Qcs\cup\Qc}}$ not to be equal to zero because
  $\Qcs\cup\Qc$ is composed of $k+b$ elements. This implies that \eqref{eq:defthetapartial}, \eqref{eq:4dimensions} and \eqref{eq:lem5}  are always well-defined, as their denominators are nonzero. \\

%Indeed, for any $\v\in\ker_0 (\A)$, we have
%\begin{align}
%\label{eq:kerA00}
%\A_{\Qcs\cup\Qc}\v_{\Qcs\cup\Qc}  
%%+ \A_{\Qc}\v_{\Qc} 
%+ \A_{\overline{\Qcs\cup\Qc}}\v_{\overline{\Qcs\cup\Qc}}=\mathbf{0}_m.
%\end{align}
% Now, $\v_{\overline{\Qcs\cup\Qc}}=\zerob$ would imply that the $k+b$
%dictionary columns indexed by $\Qcs\cup\Qc$ are linearly dependent according to 
%\eqref{eq:kerA00} and because $\v\ne\zerob$. This contradicts our assumption 
%$\spark(\A)>k+b$. Thus, we have $\v_{\overline{\Qcs\cup\Qc}}\ne\zerob$.\\

\begin{IEEEproof}\textit{(Lemma \ref{lem:orderingNSP})}
As an initial remark, let us mention that, for any $\v\in\ker_0(\A)$, a couple $(\Qcs,\Qc)$ maximizing the right-hand side of \eqref{eq:4dimensions} should be such that $\v_{\Qcs\backslash\Qc}$ (resp. $\v_{\overline{\Qcs\cup\Qc}}$) collects the elements of $\v$ with the largest (resp. smallest) amplitudes, because $t\mapsto t^p$ is an increasing function on $\R^+$. In the rest of the proof, we will therefore assume that $\Qcs$ and $\Qc$  satisfy this requirement. 
%  \remCS{conflit de notation entre le $\overline{\v}$ et
%    $\overline{\Qc}$... je pense en fait que cette notation
%    $\overline{\v}$ n'est pas necessaire et fais une proposition
%    sans la notation $\overline{\v}$:}  \remCS{Il manque une
%    explication (peut etre aussi dans la section VII) sur le fait que
%    d'apres la definition \eqref{eq:4dimensions}, pour $\v$ fixe, un
%    couple $(\Qcs,\Qc)$ qui maximise le terme de droite de
%    \eqref{eq:4dimensions} correspond forcement aux indices de plus
%    grandes ($\Qcs\backslash\Qc$) et de plus faibles amplitudes
%    ($\overline{\Qcs\cup\Qc}$) des elements de $\v$, du fait de la
%    croissante de $t\mapsto t^p$ sur $\Rbb^+$. La suite du
%    raisonnement est liee a un tel couple optimal ($\Qcs,\Qc$):}
%  \suppCS{Let $\bar{\v}\in\R^{n}$ be a vector constructed by sorting
%    the elements of $\v$ in decreasing order of magnitude. } Let
%  moreover $\w\in\R^{n-(g+b)}$ be defined as $\w=[\bar{v}_1, \ldots,
%  \bar{v}_{k-g}, \bar{v}_{g+b+1}, \ldots, \bar{v}_n]^T$
%  \remCS{remplacer par
%    $\w^T=[\v_{\Qc^\star\backslash\Qc}^T,\v_{\overline{\Qc^\star\cup\Qc}}^T]$}.

    Let $\w^T\triangleq[\v_{\Qc^\star\backslash\Qc}^T,\v_{\overline{\Qc^\star\cup\Qc}}^T]$. 
    Taking our initial remark into account, $\theta_p(k,g,b,\v)$ can be expressed as
%\begin{align}
%\theta_p(k,g,b,\v)
%= \frac{\sum_{i=1}^{k-g}\vert \bar{v}_i \vert^p}{\sum_{i=g+b+1}^{n}\vert \bar{v}_i \vert^p} 
%= \frac{\sum_{i=1}^{k-g}\vert w_i \vert^p}{\sum_{i=k-g+1}^{n-g-b}\vert w_i \vert^p}.
%\end{align}
%\remCS{a remplacer par:}
\begin{align}
\theta_p(k,g,b,\v)
= \frac{\|\v_{\Qc^\star\backslash\Qc}\|_p^p}{\|\v_{\overline{\Qc^\star\cup\Qc}}\|^p_p} 
= \frac{\sum_{i=1}^{k-g}\vert w_i \vert^p}{\|\w\|_p^p-\sum_{i=1}^{k-g}\vert w_i \vert^p}.
\end{align}
Showing \eqref{lem:ordering1} is therefore equivalent to proving that
%\begin{align}
%\frac{\sum_{i=1}^{k-g}\vert w_i \vert^q}{\sum_{i=k-g+1}^{n-g-b}\vert w_i \vert^q}\leq \frac{\sum_{i=1}^{k-g}\vert w_i \vert^p}{\sum_{i=k-g+1}^{n-g-b}\vert w_i \vert^p}\qquad \mbox{for $q<p$}.\label{app:lem5_1}
%\end{align}
%\remCS{a remplacer par:}
\begin{align}
\frac{\sum_{i=1}^{k-g}\vert w_i \vert^q}{\|\w\|_q^q-\sum_{i=1}^{k-g}\vert w_i \vert^q}
\leq \frac{\sum_{i=1}^{k-g}\vert w_i \vert^p}{\|\w\|_p^p-\sum_{i=1}^{k-g}\vert w_i \vert^p}.
\end{align}
which can also be rewritten as
\begin{align}
\frac{\sum_{i=1}^{k-g}\vert w_i \vert^q}{\|\w\|^q_q}\leq \frac{\sum_{i=1}^{k-g}\vert w_i \vert^p}{\|\w\|^p_p}\qquad \mbox{for $q<p$}.\label{app:lem5_2}
\end{align}
Now, in \cite[Th. 5]{Gribonval2007Highly}, it is proved that
\eqref{app:lem5_2} holds for any vector $\w$ whose $k-g$ first
  elements have the largest magnitudes. Observing that $\w$ satisfies the
latter condition, we obtain the result.

\end{IEEEproof}

\begin{IEEEproof}\textit{(Lemma \ref{lem:orderingNSPERClem})}
For any $\v\in\ker_0 (\A)$, we have
\begin{align}
%\label{eq:kerA0}
\A_{\Qcs\backslash\Qc}\v_{\Qcs\backslash\Qc} = 
-\A_{\Qc}\v_{\Qc} - \A_{\overline{\Qcs\cup\Qc}}\v_{\overline{\Qcs\cup\Qc}}.
\end{align}
Applying the orthogonal projector onto $\spa(\A_\Qc)^\perp$ to both sides, we obtain
\begin{align}
\tA_{\Qcs\backslash\Qc}\v_{\Qcs\backslash\Qc} =  - \tA_{\overline{\Qcs\cup\Qc}}\v_{\overline{\Qcs\cup\Qc}}.
\end{align}
Let us note that $\A_{\Qcs\cup\Qc}$ is full-rank by hypothesis and, by virtue of \cite[Corollary~3]{Soussen2013Joint}, $\tA_{\Qcs\backslash\Qc}$ is therefore also a full-rank matrix. This leads to
\begin{align}
\v_{\Qcs\backslash\Qc} =  - \tA_{\Qcs\backslash\Qc}^\dag\tA_{\overline{\Qcs\cup\Qc}}\v_{\overline{\Qcs\cup\Qc}}.
\end{align}
Taking the $\ell_1$ norm of both sides and using the definition of $\ell_1$ induced norms, we have
\begin{align}
\frac{\| \v_{\Qcs\backslash\Qc}\|_1}{\|\v_{\overline{\Qcs\cup\Qc}}\|_1} \leq  \max_{i\notin{\Qcs\cup\Qc}} \| \tA_{\Qcs\backslash\Qc}^\dag\ta_i\|_{1}. \label{eq:prooflemma6}
% \max_{i\notin{\Qcs}}\| \tA_{\Qcs\backslash\Qc}^\dag\a_i\|_1
\end{align}
%Let us mention that the left-hand side of \eqref{eq:prooflemma6} is always properly defined.
% Indeed, $\v_{\overline{\Qcs\cup\Qc}}=\zerob$ would imply that the ($k+b$)  
%dictionary columns indexed by $\Qcs\cup\Qc$ are linearly dependent according to 
%\eqref{eq:kerA0} and because $\v\ne\zerob$. This contradicts our assumption 
%$\spark(\A)>k+b$. Thus, we have $\v_{\overline{\Qcs\cup\Qc}}\ne\zerob$.
 The result then follows from the fact that $\ta_i=\mathbf{0}_m$ $\forall i\in\Qc$.
\end{IEEEproof}
\vspace{0.4cm}

\section{Proof of the results of section \ref{sec:NC}}\label{app:NC}

In this appendix, we provide a proof of Lemmas \ref{lem:Ctilde} and \ref{lem:SCOMP_NC}.
We use the notation $\Rc$ instead of $\Qc$ 
whenever the current
  support reached by Oxx has a cardinality that differs from
  $g+b$. The notation $\Rc$ is introduced to avoid any confusion, 
  since in the rest of the paper, we always have $\vert\Qc\vert=g+b$.

%to denote the
%  current support. This change of notation is done to avoid confusion:
%  in the rest of the paper, we have $\vert\Qc\vert=g+b$ whereas in this
%  appendix, the support cardinality may differ.

We first need to prove the following technical lemma:
\begin{lemma} \label{lem:symproblem} Let $\A$ be defined as in
  \eqref{eq:defDico}. Then, we have for all $\Rc$ with $\vert
  \Rc \vert <2k-g+b$ and $i, j\notin \Rc$, $i\neq j$:
%Then, we have $\forall\, \Qc$ with $\vert \Qc \vert = l<2k-l$ and $h,i, j\notin \Qc$:
%\begin{align}
%\stdscal{\ta_h,\ta_i} &= \stdscal{\ta_h,\ta_j}<0\quad  \mbox{$i\neq h, j\neq h $},\label{lem:c1}\\
%\| \ta_i \| &= \| \ta_j \|.
%\end{align}
\begin{align}
\stdscal{\ta_i^{\Rc},\ta_j^{\Rc}}&=-\mu-\mu^2\mathbf{1}_{\vert\Rc\vert}^T (\A_\Rc^T \A_\Rc)^{-1} \mathbf{1}_{\vert\Rc\vert},\label{lem:c1}\\
\| \ta_i^{\Rc}\|^2&= 1-\mu^2\mathbf{1}_{\vert\Rc\vert}^T (\A_\Rc^T \A_\Rc)^{-1} \mathbf{1}_{\vert\Rc\vert}.\label{lem:c1b}
\end{align}
\end{lemma}

\begin{IEEEproof} First recall that $\spark(\A)=2k-g+b$ (see section
  \ref{sec:NC}). Therefore, $\A_\Rc$ is full rank when $\vert
  \Rc\vert<2k-g+b$ and $\ta_i^{\Rc}$ reads
\begin{align}
\ta_i^{\Rc} = 
\mathbf{P}_\Rc^\bot 
\a_i= \a_i - \mathbf{P}_{\Rc}\a_i = \a_i - \A_{\Rc} (\A_{\Rc}^T \A_{\Rc})^{-1} \A_{\Rc}^T \a_i.
\end{align}
Using this expression, we have
\begin{align}
\stdscal{\ta_i^{\Rc},\ta_j^{\Rc}}&=\stdscal{\a_i,\a_j}-\a_i^T\A_{\Rc} (\A_{\Rc}^T \A_{\Rc})^{-1} \A_{\Rc}^T \a_j,\\
\| \ta_i^{\Rc}\|^2&= 1- \a_i^T\A_{\Rc} (\A_{\Rc}^T \A_{\Rc})^{-1} \A_{\Rc}^T \a_i.
\end{align}
Taking into account that the inner product between any pair of atoms is equal to $-\mu$ by definition of $\G=\A^T\A$, we obtain the result. %these expressions become
%\begin{align}
%\stdscal{\ta_i,\ta_j}&=-\mu-\mu^2\mathbf{1}_{l}^T (\A_\Qc^T \A_\Qc)^{-1} \mathbf{1}_{l},\\
%\| \ta_i\|^2&= 1-\mu^2\mathbf{1}_{l}^T (\A_\Qc^T \A_\Qc)^{-1} \mathbf{1}_{l}.
%\end{align}

\end{IEEEproof}

\begin{IEEEproof}\emph{(Lemma~\ref{lem:Ctilde})}
 % \addCS{Since $\spark(\A)=2k-l\geq  2(l+1)-l=l+2$, $\A_{\Qc\cup\{i\}}$ is full rank for $i\notin\Qc$,  so $\tc_i^{\Qc}\ne\textbf{0}$ according  to~\cite[Corollary~3]{Soussen2013Joint}.}
% 
Using Lemma \ref{lem:symproblem} for $\vert\Rc\vert=g+b$,
    we notice that 
    $\tC_{\backslash\Qc}^{\Qc}=\beta
    \tA_{\backslash\Qc}^{\Qc}$ for
    some $\beta>0$ since $\| \ta_i^{\Qc} \|$ does not depend on $i$. Moreover, $\ta_i^{\Qc}\neq \mathbf{0}_m$ (and therefore $\tc_i^{\Qc}\neq \mathbf{0}_m$) since
    $\spark(\A)=2k-g+b>g+b+1$, which implies that $\A_{\Qc\cup\{i\}}$ is full-rank and, in turn, that $\ta_i^\Qc\neq\mathbf{0}_m$. Defining $\v\triangleq\mathbf{1}_{2k-2g}$,
    we obtain
  \begin{align}
    \tC_{\backslash\Qc}\v
    &= \beta \tA_{\backslash\Qc}\v\\
    & =\beta\tA \mathbf{1}_{2k-g+b}= \beta \proj\A\mathbf{1}_{2k-g+b} = \mathbf{0}_{2k-g+b-1} \label{eq:detailsNC1}
  \end{align}
  since $ \mathbf{1}_{2k-g+b}$ belongs to the null-space of
    $\A$.

Let us partition the elements of $\v=\mathbf{1}_{2k-2g}$ into
    two subsets $\Qc_1\cup\Qc_2$ with $\Qc_1\cap \Qc_2=\emptyset$ and
    $\vert \Qc_1\vert=\vert \Qc_2\vert=k-g$, and define
    $\tilde{\y}\triangleq\tC_{\Qc_1\backslash\Qc}\mathbf{1}_{k-g}$. According
    to~\eqref{eq:detailsNC1}, $\tilde{\y}$ rereads
    $-\tC_{\Qc_2\backslash\Qc}\mathbf{1}_{k-g}$, therefore $\tilde{\y}$ has
    two $(k-g)$-sparse representations with disjoint supports in
    $\tC_{\backslash\Qc}$. 

\end{IEEEproof}

\begin{IEEEproof} (\emph{Lemma \ref{lem:SCOMP_NC}})
Let us first recall that $b$ is set to 0 in this lemma.
  We prove a result slightly more general than the statement of Lemma
  \ref{lem:SCOMP_NC}: for the dictionary defined as in
  \eqref{eq:defDico}, any subset $\Rc$ with
    $p\triangleq\vert\Rc \vert\leq 2k-g-2$ can be reached by Oxx.
  Lemma \ref{lem:SCOMP_NC} corresponds to the case $p=g$
  ($p\leq 2k-g-2$ is always satisfied as long as $g<k$).

  The result is true for OLS by virtue of
  \cite[Lemma~3]{Soussen2013Joint} which states that any subset
  $\Rc$ of an \emph{arbitrary} dictionary $\A$ is reachable as long as
  $\vert \Rc\vert\leq \spark(\A)-2$. In particular, the latter
  condition is verified by the dictionary $\A$ and the subset $\Rc$
  considered here since $\spark(\A)=2k-g$ and $\vert\Rc \vert\leq
  2k-g-2$ by hypothesis.

  We prove hereafter that the result is also true for OMP. Without
  loss of generality, we assume that the elements of $\Rc$ correspond
  to the first $p$ atoms of $\A$ (the analysis
    performed hereafter remains valid for any other support $\Rc$ of
    cardinality $p$ since the content of the Gram matrix
    $\A_\Rc^T\A_\Rc$ is constant whatever the support $\Rc$:
    see~\eqref{eq:prod_scal_atomes}).
  % From the definition of
  % reachability (see Definition \ref{def:reachability}) we want
  % therefore to show that there exists $\y\in\spa\{\a_1, \ldots,
  % \a_q\}$ such that OMP with $\y$ as input selects atoms in
  % $\Qc=\{1,\ldots,q\}$ during the first $q$ iterations.
  %
  % 
  For arbitrary values of $\varepsilon_2,\ldots, \varepsilon_{p}>0$,
  we define the following recursive construction:
  \begin{itemize}
    \item $\y_1=\a_1$,
    \item $\y_{p+1}=\y_{p}+\varepsilon_{p+1}\a_{p+1}$ 
  \end{itemize}
  ($\y_{p+1}$ implicitly depends on
  $\varepsilon_2,\ldots,\varepsilon_{p+1}$).  We show by recursion
  that for all $p\in\{1,\ldots,2k-g-2\}$, there exist
  $\varepsilon_2,\ldots,\varepsilon_{p}>0$ such that OMP with the
  dictionary defined as in~\eqref{eq:defDico} and $\y_{p}$ as input
  successively selects $\a_1,\ldots,\a_{p}$ during the first $p$
  iterations. In particular, the selection
  rule~\eqref{eq:atomselection2} always yields a unique maximum.

  The statement is obviously true for $\y_1=\a_1$. Assume that it is
  true for $\y_{p}$ ($p<2k-g-2$) with some $\varepsilon_2,\ldots,
  \varepsilon_{p}>0$ (these parameters will remain fixed in the
  following). According to \cite[Lemma 15]{Soussen2013Joint}, there
  exists $\varepsilon_{p+1}>0$ such that OMP with
  $\y_{p+1}=\y_{p}+\varepsilon_{p+1}\a_{p+1}$ as input selects the same
  atoms as with $\y_{p}$ during the first $p$ iterations, \ie
  $\a_1,\ldots,\a_{p}$ are successively chosen. At iteration $p$, the
  current active set reads $\Rc=\{1,\ldots,p\}$ and the corresponding
  residual takes the form
  \begin{align}
    \r^{\Rc}= \varepsilon_{p+1} \ta_{p+1}^{\Rc}.
  \end{align}
  Thus, $\a_{p+1}$ is chosen at iteration $p+1$ if and only if
  \begin{align}\label{eq:goodselectionOMP}
    \vert\stdscal{\ta_i^{\Rc},\ta_{p+1}^{\Rc}}\vert< 
\| \ta_{p+1}^{\Rc} \|^2 \qquad \forall \, i\neq p+1. 
  \end{align}

  Now, $\vert \Rc\vert=p<2k-g$ by hypothesis, then Lemma
  \ref{lem:symproblem} applies (we remind the reader that we assume that $b=0$). Using \eqref{lem:c1}-\eqref{lem:c1b},
  it is easy to see that \eqref{eq:goodselectionOMP} is equivalent to
  \begin{align}\label{eq:cnsgoodselectionOMP}
    \mu + 2 \mu^2 \mathbf{1}_{p}^T (\A_{\Rc}^T \A_{\Rc})^{-1} \mathbf{1}_{p} < 1.
  \end{align}
  Since $\mu=\frac{1}{2k-g-1}<\frac{1}{p+1}<\frac{1}{p-1}$, we have
  $1-(p-1)\mu>0$. Then, \cite[Lemma~2.3]{Tropp2004Greed} and
    $\|\mathbf{1}_{p}\|^2=p$ yield:
  \begin{align}
    \mathbf{1}_{p}^T (\A_{\Rc}^T \A_{\Rc})^{-1} \mathbf{1}_{p}\leq \frac{p}{1-(p-1)\mu}. 
  \end{align}
Using the majoration $\mu<1/(p+1)$, it follows that:
  \begin{align}
    \mu + 2 \mu^2 \mathbf{1}_{p}^T (\A_{\Rc}^T \A_{\Rc})^{-1}
    \mathbf{1}_{p} &\leq
    \mu\left (1+  \frac{2 \mu p}{1-(p-1)\mu}\right )\\
    &= \mu\left (\frac{1+(p+1)\mu}{1-(p-1)\mu}\right )\\
    &< \frac{1}{p+1}\left (\frac{2}{1-\frac{p-1}{p+1}}\right )=1
  \end{align}
which proves that the
    condition~\eqref{eq:cnsgoodselectionOMP}, and
    then~\eqref{eq:goodselectionOMP} is met.
  % \begin{align}
  %   \mu<\frac{1}{p+1}. 
  % \end{align}
  % or equivalently
  % \begin{align}
  %   p<2k-l-2. 
  % \end{align}
OMP therefore recovers the subset
    $\Rc\cup\{p+1\}=\stdacc{1,\ldots,p+1}$.

\end{IEEEproof}

\bibliographystyle{IEEEbib}
%\bibliography{../../../References/bibliography_new.bib}
\bibliography{group-15302_mod}

\end{document}